\documentclass[fleqn,usenatbib]{mnras}
\usepackage{newtxtext,newtxmath}

\usepackage[T1]{fontenc}

\DeclareRobustCommand{\VAN}[3]{#2}
\let\VANthebibliography\thebibliography
\def\thebibliography{\DeclareRobustCommand{\VAN}[3]{##3}\VANthebibliography}

\usepackage{graphicx}	
\usepackage{amsmath}	
\usepackage{xcolor}
\usepackage{subcaption}
\usepackage{caption}
\usepackage{academicons}
\usepackage{hyperref}
\usepackage{orcidlink}
\usepackage[normalem]{ulem}

\newcommand{\lcdm}{$\Lambda\rm{CDM}$} 

\defcitealias{Macpherson_Heinesen_2021}{MH21}
\defcitealias{dhawan2022quadrupole}{D22}

\title[Constraints on a quadrupolar expansion in Pantheon+]{Potential signature of a quadrupolar Hubble expansion in Pantheon+ supernovae}

\author[Jessica A. Cowell et al.]{
Jessica A. Cowell \orcidlink{0000-0002-7228-6621},$^{1,2, 3}$\thanks{E-mail: jessica.cowell@physics.ox.ac.uk(JAC) , 
}
Suhail Dhawan,$^{1}$
Hayley J. Macpherson,$^{4,5,6}$
\\
$^{1}$Institute of Astronomy and Kavli Institute for Cosmology, University of Cambridge, Madingley Road, Cambridge CB3 0HA, UK\\
$^{2}$ Department of Physics, University of Oxford, Denys Wilkinson Building, Keble Road, Oxford OX1 3RH, UK\\
$^{3}$ Kavli Institute for the Physics and Mathematics of the Universe (IMPU), the University of Tokyo, Kashiwa, Chiba, 277-8582, Japan\\
$^{4}$Department of Applied Mathematics and Theoretical Physics, University of Cambridge, Cambridge CB3 0WA, UK \\
$^{5}$Kavli Institute for Cosmological Physics, The University of Chicago, 5640 South Ellis Avenue, Chicago, Illinois 60637, USA \\
$^{6}$NASA Einstein Fellow
}

\date{Accepted XXX. Received YYY; in original form ZZZ}

\pubyear{2022}

\begin{document}
\label{firstpage}
\pagerange{\pageref{firstpage}--\pageref{lastpage}}
\maketitle

\begin{abstract}
The assumption of isotropy --- that the Universe looks the same in all directions on large scales --- is fundamental to the standard cosmological model. This model forms the building blocks of essentially all of our cosmological knowledge to date. It is therefore critical to empirically test in which regimes its core assumptions hold. Anisotropies in the cosmic expansion are expected on small scales due to nonlinear structures in the late Universe, however, the extent to which these anisotropies might impact our low-redshift observations remains to be fully tested. 
In this paper, we use fully general relativistic simulations to calculate the expected local anisotropic expansion and identify the dominant multipoles in cosmological parameters to be the quadrupole in the Hubble parameter and the dipole in the deceleration parameter. We constrain these multipoles simultaneously in the new Pantheon+ supernova compilation. The fiducial analysis is done  in the rest frame of the CMB with peculiar velocity corrections.
Under the fiducial range of redshifts in the Hubble flow sample, we find a $\sim 2\sigma$ deviation from isotropy. We constrain the  eigenvalues of the quadrupole in the Hubble parameter to be $\lambda_1 =0.021\pm{ 0.011}$ and $ {\lambda_2= 3.15\times 10^{-5}}\pm 0.012$ and place a $1\sigma$ upper limit on its amplitude of $2.88\%$. We find 
no significant dipole in the deceleration parameter,  finding constraints of  $q_{\rm dip} =  4.5^{+1.9}_{-5.4}$. However, in the rest frame of the CMB without corrections, we find $ q_{ \rm dip} = 9.6^{+4.0}_{-6.9}$, a $>2\sigma$ positive amplitude. We also investigate the impact of these anisotropies on the Hubble tension. We find a maximal shift of $0.30$ km s$^{-1}$ Mpc$^{-1}$ in the monopole of the Hubble parameter and conclude that local anisotropies are unlikely to fully explain the observed tension.

\end{abstract}




\section{Introduction}

The Lambda {cold dark matter} model (\lcdm) is generally accepted as the current standard model of cosmology.
Over the years, \lcdm\ has amassed overwhelming agreement from many {cosmological} measurements, notably measurements from the {cosmic microwave background (CMB)} polarization, temperature, and lensing \citep{Planck2018}, 
the {clustering} of galaxies at large scales \citep[e.g.][]{DESY3_cosmo}, and the measurements of the acceleration of the expansion of the Universe \citep{Riess1998,Perlmutter1999}.
Despite its many successes, 
some disagreements between \lcdm\ predictions and observations {are coming to light as measurements get more precise}. Perhaps the most notable is the ``Hubble tension'' \citep{Riess_2016}{:} the disagreement between inferences of the Hubble parameter {at redshift zero}, $H_0$, from Cepheid-calibrated supernovae {distances} 
and predictions from the CMB {which assume \lcdm}. 
{Many works have considered a wealth of} possible phenomenological or systematic sources of the tension 
\citep[see, e.g.][]{Di_Valentino_2021, Efstathiou_2021, Mortsell2018}{, however, no one solution has yet been widely accepted. Aside from the Hubble tension, there are other disagreements with 
\lcdm\ with varying significance}, 
see, e.g., \citet{abdalla2022cosmology} {and \citet{Aluri2022}} for recent reviews.

For such a long-standing model, as our data become more precise it is imperative {that} we continue to test the validity of the assumptions {upon which the model was originally built. Departures from these assumptions may only become observable once our precision passes a certain threshold, thus, continuous testing is required to ensure both precision and accuracy in cosmology. } 
{While the CMB radiation we observe is largely isotropic --- after removing the dipole, which still leaves several anomalies \citep[see, e.g.][]{Schwarz:2016} --- tests of isotropy in the late Universe are in disagreement \citep[e.g.][]{Ripa2017,Javanmardi2017,Alonso_2015,Gibelyou_2012}. }
In particular, at what point can we assume a transition to global isotropy from the anisotropic local Universe, 
{where effects such as local bulk motions will dominate}? {Recent studies have called the assumption of isotropy on small scales into question \citep[e.g.][]{Colin_2011}.}

A cornerstone of \lcdm\ is the assumption of a flat Friedmann–Lema\^itre–Robertson–Walker (FLRW) {space-time} metric. {These models assume \textit{exact} homogeneity and isotropy of space-time, however, their use is motivated by observations of the transition to \textit{statistical} homogeneity and isotropy at large scales \citep[e.g.][]{Scrimgeour:2012,Hogg:2005}. } 
{The FLRW assumption has allowed us to extract cosmological information from our observations even in early cosmology when data sets were limited.} 
However, in the coming years, the amount of data is expected to drastically {increase} \citep{LSST_SRD, Hounsell2018,scolnic2019generation, Simons_Observatory_2019}, which will allow us to 
{critically} investigate the {realm of} validity of the 
assumptions at the core of \lcdm.

{ There is recent debate around whether the presence of any observed anisotropy is still consistent with \lcdm\ once 
{we consider local peculiar velocity} effects in the local Universe. 
{In particular, recent} discussion {has revolved} around the CMB dipole \citep{2018CMB, 2003WMAP.}. 
{This dipole is commonly} 
attributed to the motion of our local galactic group (LG) relative to the CMB, 
{caused by gravitational} attraction to a nearby over-density 
{\citep[see, e.g.][]{Nusser_Davis_2011}}. 
Many studies have investigated whether the motion of the LG is consistent with the {measured} CMB dipole direction. While some find 
agreement \citep[e.g.][]{Feindt2013, Appleby_2015}, other studies using Type Ia supernovae (SNe) claim to find no bulk flow \citep[e.g.][]{Huterer_2015}. 
More generally, there is still heavy debate around the presence of anisotropy in low redshift data. While some studies claim to find consistency with {\lcdm\ } 
\citep[e.g.][]{Gibelyou_2012, Alonso_2015, chang_2018}, recent 
{work by \citet{2022HAYLEY} shows that we expect anisotropies} 
in the distance-redshift relation for low-redshift data sourcing from local differential expansion. {Any universe with structure will contain such anisotropies, so what remains is to determine their significance in our cosmological data. } 
{Some works} using the hemispherical comparison method \citep{2016Bengaly, Cai_2012} 
{or} the 
cosmographic method 
\citep[][see also Section~\ref{sec:cosmo_distances}]{Sarkar2019, Wang_Wang_2014} find {a significant dipole} in the deceleration parameter using SNe and gamma ray burst data. Furthermore, \citet{Secrest_2021} found a large dipole in the angular distribution of quasars and \citet{Bolejko2016} found dipolar and quadrupolar anisotropies in the Hubble expansion. Moreover \citet{Rameez_2018, Kalbouneh:2022tfw} found anisotropies using galaxy catalogues. 
Meanwhile, 
{some studies have found consistency} 
with \lcdm\ using both the hemispherical method \citep{Kalus2013, zhao2019} 
{and other dipole fitting methods}
\citep[e.g.][]{Rubin_2020, Andrade2018, rahman2021new, soltis2019, Alonso_2015, Gibelyou_2012}.}

{Many of these} anisotropy studies 
are independent of a particular cosmological model {through their use of a cosmographic expansion of the luminosity distance \citep[see][and Section~\ref{sec:cosmo_distances}]{Visser_2004}. Typically, anisotropies are added on top of the background FLRW cosmographic expansion. } 
In this work we will use a physically-motivated approach, 
{via our use of} the novel {generalised cosmographic expansion} 
presented in \citet{Heinesen_2021}. This framework is 
independent {of any form of the metric tensor or field equations,} which allows 
for a truly model independent analysis. In practise, such an analysis is difficult, owing to the vastly increased number of independent degrees of freedom (DOFs) of the general formalism {with respect to the FLRW framework}. Some works have tried to reduce the DOFs either by considering realistic physical approximations \citep{Heinesen_Macpherson_2021_Hubble_flow} or by analysing the framework within numerical relativity (NR) simulations \citep{Macpherson_Heinesen_2021}.

We extend on the work of \citet{Macpherson_Heinesen_2021} (hereafter referred to as \citetalias{Macpherson_Heinesen_2021}) by performing a quantitative analysis {of their same data} 
to determine which anisotropic signatures we expect to be dominant. {Then, we} constrain the {dominant anisotropies we find} in the {new} Pantheon+ {SNe} data set \citep{scolnic2021pantheon+}. 
{In doing this, we} improve the recent constraints from \citet{dhawan2022quadrupole} (henceforth referred to as \citetalias{dhawan2022quadrupole}) --- where the authors found no significant quadrupole in the Hubble parameter {using the Pantheon data set} \citep{pantheon_2018}.

\section{Cosmological Distances}\label{sec:cosmo_distances}

{In practice, when measuring distances in cosmology we use the distance luminosity relation, a relation between the distance of an object {and its redshift, which may assume some specific} 
cosmological model {(i.e., some expansion history)}. 
{However,} at low redshift, we can free ourselves from these constraints 
using the cosmographic approach. 
{Many} surveys measuring distance to astronomical objects at low redshift ($z\ll 1$) 
{will use a Taylor expansion of} the luminosity-distance relation. 
{Such an expansion can then be used in conjunction with observational data to infer cosmological parameters without assuming a specific expansion history.}}

{In Section~\ref{sec: FLRW_cosmography} below, we briefly introduce the standard expansion performed within FLRW cosmologies, and in Section ~\ref{sec:general_cosmog} we briefly discuss a generalised formalism which does not assume a particular space-time metric.}

\subsection{FLRW Cosmography}
\label{sec: FLRW_cosmography}

The standard approach for many surveys measuring the distance to astronomical objects at low redshift {($z<1$)} is to Taylor expand the luminosity distance {as a function of redshift, usually truncated at third order, namely } 
\begin{equation}
d_{L}(z) = d^{(1)}_{L}z +d^{(2)}_{L}z^2+d^{(3)}_{L}z^3 + \mathcal {O}(z^4) .
\label{eq:flrw_taylor}
\end{equation}
{Within the FLRW class of models, }
the coefficients of Eq.~\eqref{eq:flrw_taylor} can be {expressed as \citep{Visser_2004}}
\begin{subequations}\label{eqs:dLFLRW_coeff}
  \begin{align}
    d^{(1)}_{L,FLRW}& \equiv \frac{1}{H_o},\\
    d^{(2)}_{L,FLRW}& \equiv \frac{1-q_o}{2H_o},\\
    d^{(3)}_{L,FLRW} & \equiv \frac{-1 + 3q_o^2 + q_o - j_o + \Omega_{k,o} }{6H_o}.
\end{align}

\end{subequations}
In the above, the standard Hubble, deceleration, jerk and curvature parameters, are defined as 
\begin{align}
\label{eq:h0}
&H \equiv \;\frac{\dot{a}}{a}, 
\quad\quad\; q \equiv\; - \frac{\ddot{a}}{aH^2},  \\
& j \equiv \frac{ \dddot{a}}{aH^3}, 
 \quad\quad \Omega_k \equiv \frac{-k}{a^2H^2}, \label{eq:omega_k}
\end{align}
respectively, where $a$ is the FLRW scale factor, $k$ is the scalar curvature of the space-time {--- taking on values $k = \pm1, 0$ ---}
{and an over-dot represents a derivative with respect to time.} {The subscript `$o$' in Eqs.~\eqref{eqs:dLFLRW_coeff} denotes that the parameters are measured at the observer position, 
i.e. at redshift $z=0$.}

The definitions {above} are explicitly dependent on the existence of an {exact} FLRW geometry {and expansion}. 
In the next section, we will summarise the recent results of \citet{Heinesen_2021} in which 
{the author derived} the {fully} model-independent {cosmographic} 
expansion of $d_L$ in $z$, making no assumptions on the {field equations or underlying} metric {of space-time}. 

\subsection{General Cosmography}
\label{sec:general_cosmog}

Series expansions of cosmological distances in the context of arbitrary space-time metrics --- i.e., removing the FLRW approximation --- have been studied for decades \citep[e.g.][]{Kristian_Sachs1966,Ellis_Nel_Maartens_Stoeger_Whitman_1985,Seitz1994LightPI,Clarkson_2011,Heinesen_2021}. 
{In this section,} we will focus on the more recent
results from \citet{Heinesen_2021}, in which the author presents a new generalised framework for cosmological data analysis outside of the FLRW metric. This is the framework upon which we will base our observational constraints presented in Section~\ref{sec:obs_results}. 

The general form of the cosmography once again follows from {the Taylor expansion} \eqref{eq:flrw_taylor}, however, we now define 
the inhomogenous and anisotropic coefficients 
as follows; 
\begin{subequations}
\begin{align}
d^{(1)}_{L} &\equiv \frac{1}{\mathfrak{H}_o},\\
d^{(2)}_{L} &\equiv \frac{1-\mathfrak{Q}_o} {2\mathfrak{H}_o},\\
d^{(3)}_{L} &\equiv \frac{-1 + 3\mathfrak{Q}_o^2 + \mathfrak{Q}_o- \mathfrak{J}_o + \mathfrak{R}_o}{6\mathfrak{H}_o}.
\end{align}
\end{subequations}
We consider a set of observers and emitters co-moving with the large-scale flow of the cosmological fluid with 4--velocity $u^\mu$.
The coefficients above 
{contain} 
parameters which appear in the same place in the $d_L(z)$ {expansion} 
as {their} FLRW counterparts, 
{however, they} have different physical interpretations. {They are thus named the \textit{effective}} Hubble, jerk, curvature and deceleration parameters, respectively, {and} are defined as; 
\begin{subequations}\label{eqs:effective_params}
\begin{align}
\label{eq:effective_hubble}
\mathfrak{H} &\equiv -\frac{1}{E}\frac{dE}{d\lambda},\\
\mathfrak{J} &\equiv \frac{1}{E^2} \frac{\frac{d^2\mathfrak{H}}{d\lambda^2}}{\mathfrak{H}^3}  -4\mathfrak{Q}-3,\label{eq:effective_jerk}\\
\mathfrak{R} &\equiv 1 + \mathfrak{Q} - \frac{1}{2E^2}\frac{k^{\mu}k^{\nu}R_{\mu\nu}}{\mathfrak{H}^2},
\label{eq:effective_curvature}\\
\mathfrak{Q} &\equiv -1 - \frac{1}{E}\frac{\frac{d\mathfrak{H}}{d\lambda}}{\mathfrak{H}^2}.
\label{eq: effective_decel}
\end{align}
\end{subequations}
In the above, $k^\mu$ is the 4--momentum of an incoming null ray, the derivative  ${d}/{d\lambda}$ is the derivative along the direction of the incoming null ray, {$E\equiv -u^\mu k_\mu$} is the photon energy function as measured by the observer, and $R_{\mu\nu}$ is the Ricci tensor
of the space-time. We would like to emphasise again that the above set of parameters depend both on the observer's location \textit{as well as} the direction of observation. 
{The latter can be better understood when writing} the effective parameters as an exact multipole {series} expansion in the direction of observation, $e^\mu$.
The expansion of the effective Hubble parameter $\mathfrak{H}$ {---} in terms of
kinematic variables of the fluid {---} is  
\begin{equation}
\mathfrak{H}(e) = \frac{1}{3}\theta - e^{\mu}a_{\mu}+e^{\mu}e^{\nu}\sigma_{\mu\nu},
\label{eq:effective hubble_multipole_expansion}
\end{equation}
where $\theta$ is the volume expansion rate, $a^\mu$ is the 4--acceleration and $\sigma_{\mu\nu}$ is the shear tensor {\citep[see, e.g.][for definitions of these fluid quantities]{Heinesen_2021}}. This is an exact representation, where the effective Hubble parameter is naturally truncated to quadrupolar order{. In the exactly} homogeneous and isotropic limit, {we have $\theta\rightarrow 3H$ and thus $\mathfrak{H}$} 
reduces to the FLRW Hubble parameter as defined in Eq.~\eqref{eq:h0}. 

The same {multipole expansion} can be done for the {effective} deceleration parameter, {namely}
\begin{multline}
    \mathfrak{Q}(e) = -1 - \frac{1}{\mathfrak{H}^2(e)} \bigg( \overset{0}{\mathfrak{q}}+e^\mu \overset{1}{\mathfrak{q}}_\mu + e^\mu e^\nu\overset{2}{\mathfrak{q}}_{\mu\nu}+ 
    e^\mu e^\nu e^\rho \overset{3}{\mathfrak{q}}_{\mu\nu\rho} +\\
e^\mu e^\nu e^\rho e^\kappa \overset{4}{\mathfrak{q}}_{\mu\nu\rho\kappa}\bigg)
     \label{eq: deceleration_multipole}  
\end{multline}
with coefficients
\begin{subequations} \label{eqs:q_coeffs}
    \begin{align}
     &\overset{0}{\mathfrak{q}} \equiv \frac{1}{3} \frac{d\theta}{d\tau} + \frac{1}{3}D_\mu a^{\mu} - \frac{2}{3}a^\mu a_\mu - \frac{2}{5} \sigma_{\mu\nu}\sigma^{\mu\nu}, \\
     &\overset{1}{\mathfrak{q}}_\mu \equiv - \frac{1}{3}D_\mu \theta  - \frac{2}{5} D_\nu \sigma^\nu_{\mu} -\frac{da_\mu}{d\tau}+a^\nu \omega_{\mu\nu}+ \frac{9}{5}a^\nu \sigma_{\mu\nu} \label{eq:q1_coefficent},  \\
      &\overset{2}{\mathfrak{q}}_{\mu\nu} \equiv \frac{d\sigma_{\mu\nu}}{d\tau} + D_{\langle _\mu}a_{\nu\rangle} + a_{\langle \mu }a_{\nu \rangle } -2\sigma_{\alpha\langle\mu \omega^\alpha_\nu\rangle} -\frac{6}{7}\sigma_{\alpha \langle \mu }\sigma^\alpha_{\nu \rangle }, \\
       &\overset{3}{\mathfrak{q}}_{\mu\nu\rho} \equiv  - D_{\langle _\mu}\sigma_{\nu\rho\rangle} -3a_{\langle \mu }\sigma_{\nu\rho \rangle } ,\label{eq:q3_coefficent}
       \\
    &\overset{4}{\mathfrak{q}}_{\mu\nu\rho\kappa} \equiv  2\sigma_{\langle \mu\nu }\sigma_{\rho\kappa \rangle },
    \end{align}
\end{subequations}
where $\langle\rangle$ implies symmetrisation over the enclosed indices, see \citet{Heinesen_2021}. {The multipole expansion Eq.~\eqref{eq: deceleration_multipole} implies that the parameter $\mathfrak{H}^2(\mathfrak{Q}+1)$ is exactly truncated at the 16-pole \citep[see][]{Heinesen_2021}. }

For $\mathfrak{R}$ and $\mathfrak{J}$ the same {expansions} can be done, 
{containing multipoles} up to {the} 16-pole and  
64-pole, {respectively,} see \citet{Heinesen_2021} for details.
{In this work, we will focus only on the multipolar expansions in the lowest order parameters, $\mathfrak{H}$ and $\mathfrak{Q}$.}

{In its full generality, this formalism contains 61 independent degrees of freedom when truncated at third order in redshift. \citet{Heinesen_Macpherson_2021_Hubble_flow} reduced this to $\sim 31$ degrees of freedom using realistic model universe assumptions, however, constraining this many degrees of freedom}
{
is difficult when working within the limitations of current {low redshift standardisable} 
data. To reduce the degrees of freedom when fitting these parameters, we 
{will first} analyse which multipole terms are expected to 
dominate, then simplify {the general cosmographic expression} to include only 
these dominant terms. 
For this analysis, we will use the numerical simulation {data} 
{presented in} \cite{Macpherson_Heinesen_2021}, 
in which the effective anisotropic parameters 
were calculated explicitly for {a set of synthetic} observers within the simulations.}
In the next section, we explain this data and present our multipole analysis determining the dominant multipoles for the effective cosmological parameters. 

\section{Anisotropy in Cosmological Simulations}

In this 
{section,} we analyse {two} {numerical relativity (NR)} {cosmological} simulations 
{to quantify potentially observable} effects generated by local {anisotropic} expansion. 
We will use the results 
{we present in this section} to {motivate} 
our choice of multipoles 
{we constrain} in SNe data {in Section~\ref{sec:obs_results}}. 

In Section~\ref{sec:sims} below we {describe} the simulation data we use and in Section~\ref{sec:sim_results} we present our detailed multipole analysis {across the} 
set of observers in these simulations.

\subsection{Numerical Relativity Simulation Data}
\label{sec:sims}

{We use calculations of the effective cosmological parameters \eqref{eqs:effective_params} from \citet{Macpherson_Heinesen_2021} (hereafter \citetalias{Macpherson_Heinesen_2021}), as calculated within a set of two NR cosmological simulations. We will use this data to determine the dominant multipoles in each effective parameter for a set of synthetic observers within the simulations. 
}

These 
simulations adopt a dust fluid approximation {(with negligible pressure, namely $P\ll \rho$)} for the matter with no dark energy \citep[see][for further details on the software and physical assumptions of the simulations]{MacphersonNR2019}.
{We are interested in how the effective cosmological parameters compare to their FLRW equivalents. }
Consequently, 
we normalise the simulation values to {the} flat, matter-dominated FLRW model {--- i.e., the} Einstein de Sitter (EdS) model. The initial conditions {of the simulations} are that of a linearly-perturbed EdS metric, with Gaussian-random initial density fluctuations mimicking the CMB. {The $z\approx 0$ snapshots of the simulations agree well with the EdS model used for the background initial data when averaged on large scales \citep[see][]{MacphersonNR2019}.}
Since the simulations contain no dark energy, they will have higher density contrasts in general compared to an equivalent \lcdm model universe. However, as discussed in \citetalias{Macpherson_Heinesen_2021}, we will see \textit{qualitatively} similar anisotropic signatures as in an equivalent simulation with $\Lambda\neq 0$, however, the amplitude of the signatures may be reduced. {Therefore, our conclusions on which multipoles are \textit{dominant} across observers should be robust. }

\begin{figure*}
    \centering
    \begin{subfigure}{.5\textwidth}
        \centering
        \includegraphics[width=\linewidth]{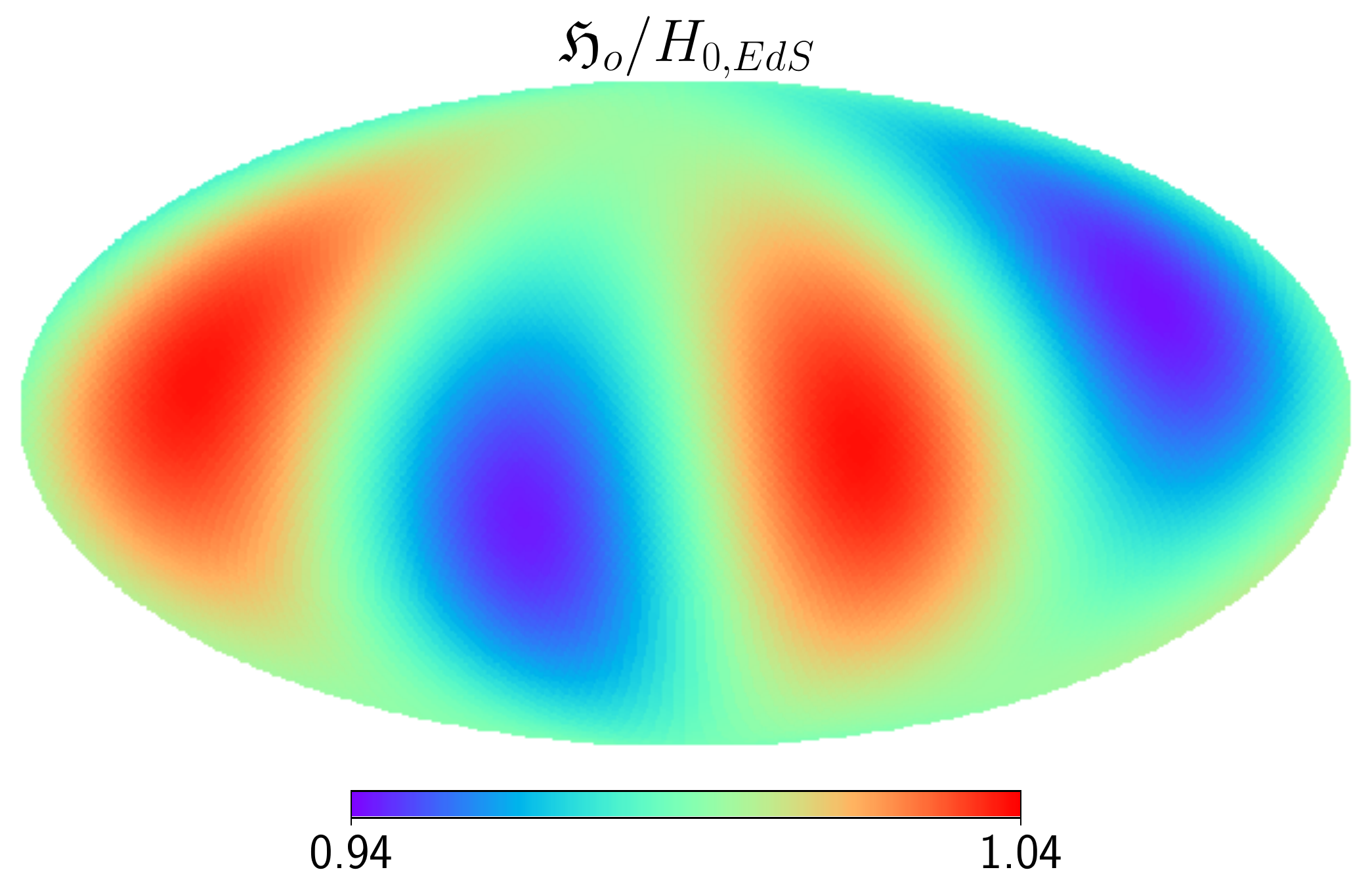}
    \end{subfigure}%
    \begin{subfigure}{.5\textwidth}
         \centering
        \includegraphics[width=\linewidth]{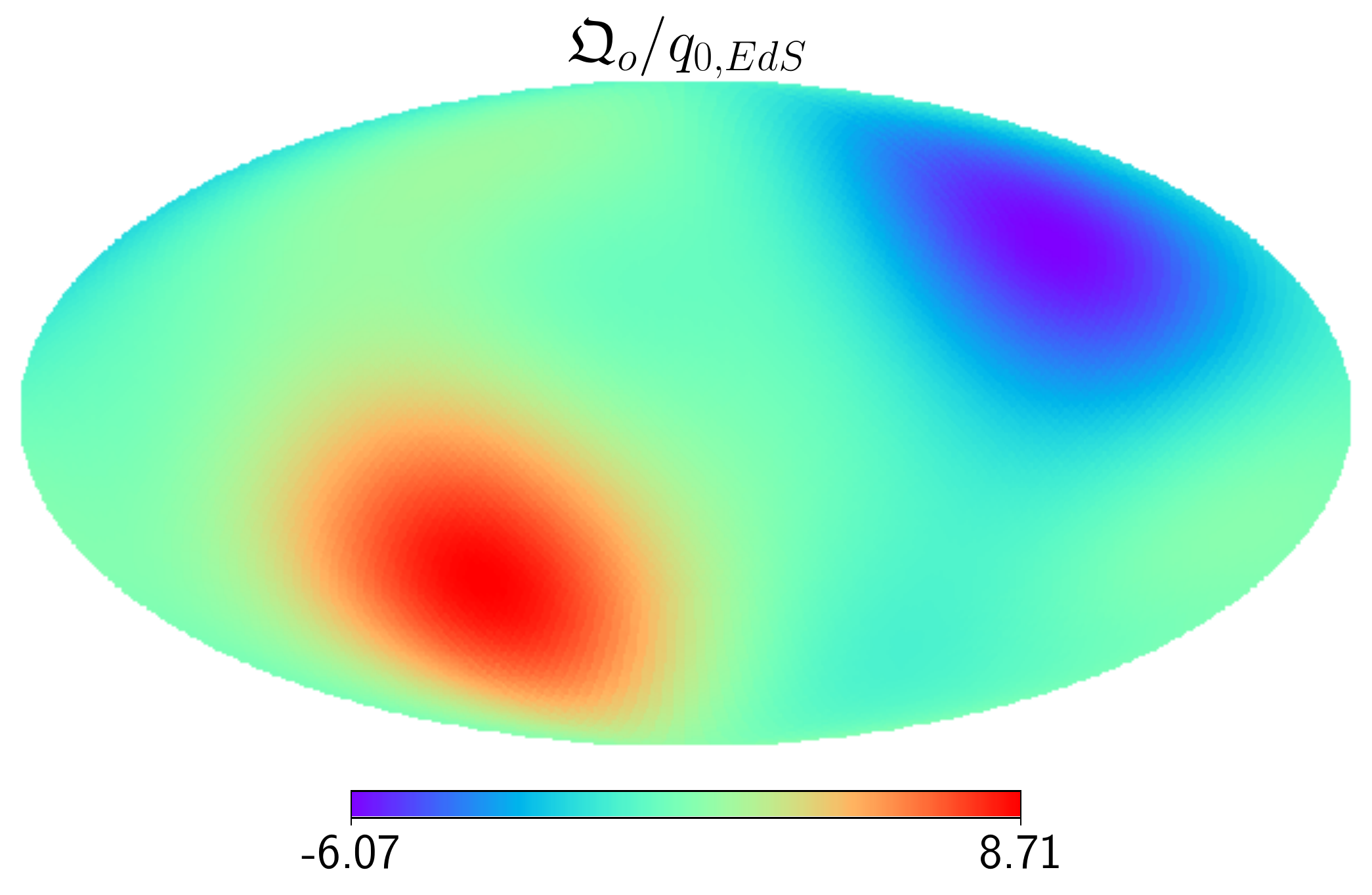}
    \end{subfigure}
    \caption{Sky maps for a single observer measured in directions of the $12\times N_{\rm side}^2$ {\tt HEALPix} pixels with $N_{\rm side} = 32$. {The simulations have physical domain length of $L=12.8 \,h^{-1}$ with a smoothing scale of $100\, h^{-1}$ Mpc. }The left panel shows the effective Hubble parameter and the right panel shows the effective deceleration parameter. Both panels are normalised by their respective EdS values.}
    \label{Fig:skymap}
\end{figure*}
We use the data of 100 observers placed in random locations throughout two simulations from \citetalias{Macpherson_Heinesen_2021}. The two simulations {both have numerical resolution $N=128$ (with the full cubic domain containing $N^3$ grid cells) but} have different {``{smoothing} 
scales''}, such that individual grid cells have lengths $100\, h^{-1}$ Mpc and $200\, h^{-1}$ Mpc, giving {domain} lengths of {$L=12.8 \,h^{-1}$} Gpc and $ L=25.6 \,h^{-1}$ Gpc{, respectively}. These smoothing scales are chosen such that small-scale non-linearities have been 
{explicitly excluded from} the simulations. 
{These chosen smoothing scales are motivated by} 
observations 
{which report} a transition to statistical homogeneity at $\approx$ 100$h^{-1}$Mpc \citep{Scrimgeour:2012wt}. {Incorporating such a smoothing scale {in these kinds of calculations} is necessary to ensure the regularity requirements of the general cosmography are satisfied in the calculations \citep[see][and also \citetalias{Macpherson_Heinesen_2021}]{Heinesen_2021}}. 

For each observer, the effective cosmographic parameters have been calculated {in the direction of}
$12\times N_{\rm side}^2$ 
{\tt HEALPix}\footnote{http://healpix.sourceforge.net} indices {with with $N_{\rm side}=32$ --- ensuring an } 

isotropic sky coverage for each observer \citep{healpy2}. 

{In} {Figure~\ref{Fig:skymap} {we show}} 
sky maps for the 
effective Hubble (left panel) and deceleration (right panel) parameters for a single observer in the NR simulation with smoothing scale {$100 h^{-1}$MPc}  (adapted from \citetalias{Macpherson_Heinesen_2021}). 
{Both parameters are normalised by their {respective} EdS
{values, namely $H_{0,{\rm EdS}}=45$ km/s/Mpc and $q_{0,{\rm EdS}}=0.5$. }
{For} this 
observer, 
by eye we can see a quadrupolar anisotropy dominating the signal {for the effective Hubble parameter --- physically} sourced from the shear tensor contribution in \eqref{eq:effective hubble_multipole_expansion}. {This is to be expected for all observers in these simulations, since they are co-moving with a dust fluid and thus the acceleration $a_\mu$ term (contributing to the dipole) is subdominant.}

For the effective deceleration parameter, 
the dipolar 
anisotropy appears to be dominant --- physically sourced by gradients in the local expansion rate. While we might expect gradient terms to dominate in the coefficients \eqref{eqs:q_coeffs}, this is not obviously the case across all observers and will be dependent on their specific location. 
{Thus, in the following section, we proceed to use this data for all observers to explicitly determine the dominant multipoles for each parameter. }


\subsection{Multipole Analysis}
\label{sec:sim_results}
 
{In this section,} we 
{work in} Fourier space where the multipole components can be easily separated. First, we use the {\tt healpy} package \citep{healpy1,healpy2} to calculate the angular power spectrum {for multipole $\ell$ as}
\begin{equation}
	\label{eq: angular power spectrum}
    {C_\ell } = \frac{1}{2 \ell +1} \sum_{m} |{{a}_{\ell m}}|^2,
\end{equation}
{using the {\tt healpy.anafast} function.}
{In the above, }$a_{\ell m}$ are the spherical harmonic coefficients to {the} spherical harmonics $Y_{\ell m }$, {which are} defined from the expansion of a band-limited function {$f$} on a sphere with angular coordinates $\theta, \phi$ as  
\begin{equation}
    \label{eq: spherical harmonics}
    f(\theta, \phi) = \sum^{\ell_{max}}_{\ell = 0}\sum_m a_{\ell m} Y_{\ell m }(\theta, \phi).
\end{equation}

We use $C_\ell$ to quantify the strength of each multipole component relative to the monopole. Specifically, for each observer, we 
{calculate} the ratio of the spherical harmonic coefficients, that is  
\begin{equation}
{\mathcal{R_{\ell}} =\sqrt{ \frac{(2\ell + 1) C_\ell}{C_{\ell=0}} } }, 
\end{equation}
for $\ell>0$ in the numerator.
\begin{figure}
    \includegraphics[width=\columnwidth]{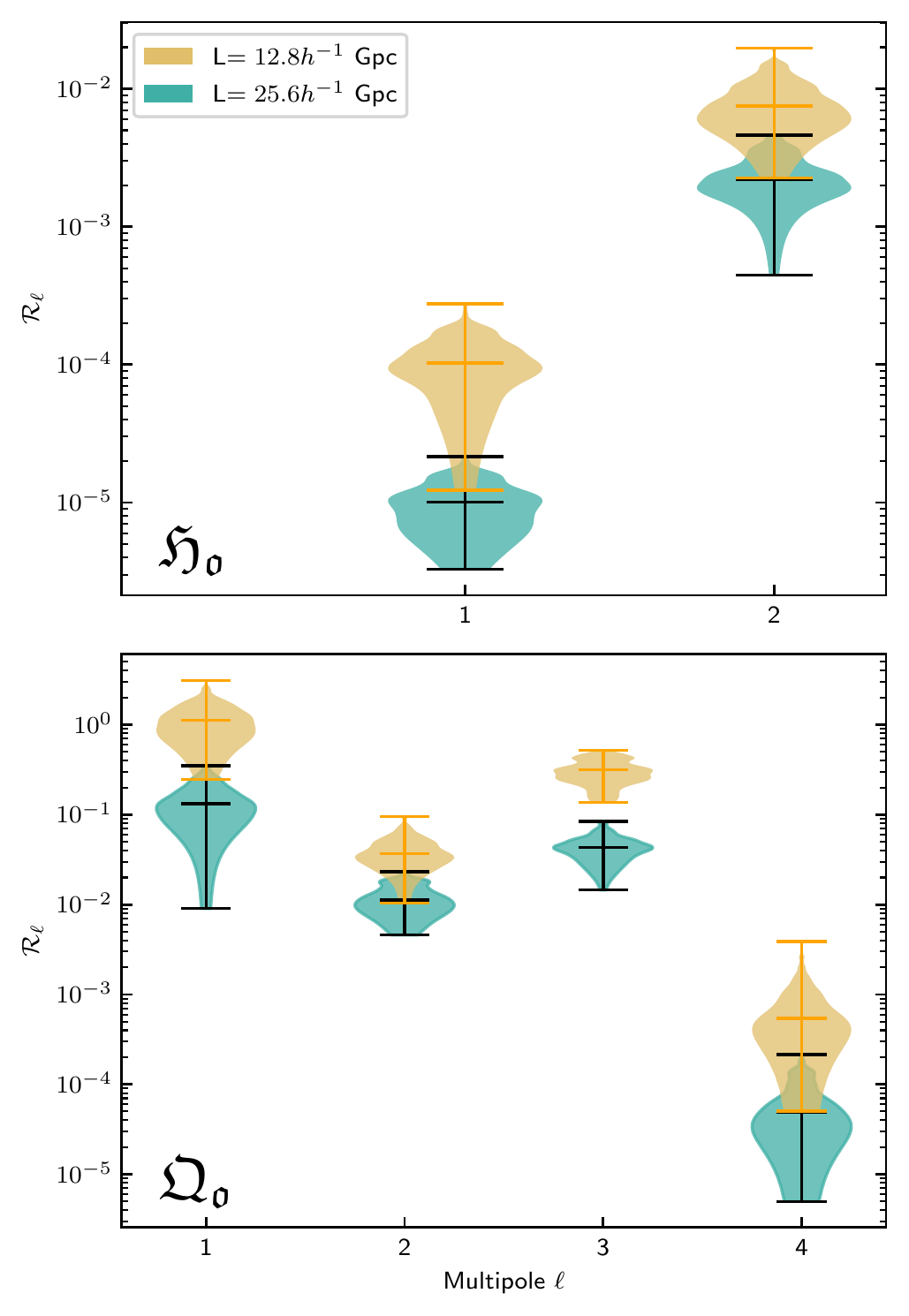}
    \caption{ 
    Relative ratios ($\mathcal{R}_\ell$) for each multipole relative to the monopole for the effective hubble ($\mathfrak{H_o}$) and deceleration ($\mathfrak{Q_o}$) parameters. The upper panel shows $\mathfrak{H_o}$, while  $\mathfrak{Q_o}$ is shown below. Shaded regions represent an empirical distribution of the data, calculated using kernel density estimation (KDE). Results for two simulations of length $12.8h^-1$Gpc and $25.6h^-1$Gpc are shown in yellow and blue. The upper limits and medians are marked on the plots with horizontal lines. Note that $\mathfrak{H_o}$ only goes up to quadrupolar order, and $\mathfrak{Q_o}$ to a 16-pole due to the exact truncations in Eq. \ref{eq:effective hubble_multipole_expansion} and \ref{eq: deceleration_multipole}}
    \label{fig:violin_plots}
\end{figure}
This approach allows us to {calculate} the relative strength of each multipole {with respect} to the isotropic {(monopole)} component. 
Since $\mathfrak{R}$ and $\mathfrak{J}$ only enter {the expansion} \eqref{eq:flrw_taylor} at third order in redshift, their contributions $d_L$ will be {difficult} to constrain with current data. {Thus, in this work}
we focus on anisotropies in the {effective} Hubble and deceleration parameters. 
{We present} the corresponding analysis for the {effective} curvature and jerk parameters in Appendix~\ref{sec: curv_jerk}.  

{In {Figure~}\ref{fig:violin_plots}} we show 
this calculation for all 100 observers for both simulations for the effective Hubble and deceleration parameters.
The shaded regions 
represent an empirical distribution of the data, calculated using kernel density estimation (KDE), with the median, upper, and lower limits displayed {as horizontal bars}. 

{We see the same {qualitative} trend} for both simulations, {however,} for the 
simulation {with $100\,h^{-1}$ Mpc smoothing scale} we see 
{generally} larger {amplitude} anisotropic components. This is expected, as the {smaller physical} resolution 
allows for more {small-scale structures to form, resulting in stronger {local} anisotropic effects}. {For the rest of this work, }we 
{consider} 
{this} simulation 
as the fiducial case{, since it contains the most realistic structure of the two. } 

For the effective Hubble parameter $\mathfrak{H}_0$, 
the dipole is 
on average {four} orders of magnitude smaller than the {monopole} term, with a median of { $\mathcal{R}_{\ell=1}(\mathfrak{H}) =1.029\times 10^{-4} $}.
However, 
{the} quadrupole 
{has} median value {$\mathcal{R}_{\ell=2}(\mathfrak{H}) = 7.52 \times 10^{-3}$. }
The effective deceleration parameter, 
$\mathfrak{Q}_0$, 
has a dipole {with amplitude} approximately {$110\%$ }{that} 
of the monopole, {namely a median of} {$\mathcal{R}_{\ell=1}(\mathfrak{Q}) = 1.12$}.{The} 
octopole of {the effective deceleration parameter has amplitude of approximately} 
{$30\%$, {or} $\mathcal{R}_{\ell=3}(\mathfrak{Q}) =0.314$. }

From these results, we {can} reduce the degrees of freedom of 
the general cosmographic expansion 
by including only the dominant terms in each effective parameter.

The simplified version of Eq. \eqref{eq: deceleration_multipole} we use for the effective deceleration parameter which includes the monopole, dipole, and octopole contributions is
\begin{equation}
    \label{eq: approx q}
    \mathfrak{Q}(e) \approx -1 - \frac{1}{\mathfrak{H}^2(e)} \bigg( \overset{0}{\mathfrak{q}}+e^\mu \overset{1}{\mathfrak{q}}_\mu + 
    e^\mu e^\nu e^\rho \overset{3}{\mathfrak{q}}_{\mu\nu\rho} \bigg{)}.
\end{equation}

There are still
many degrees of freedom even in this simplified expression. 
It will be difficult to constrain all {three multipole} components of $\mathfrak{Q}(e)$, and although the sky coverage of SNe has 
improved with Pantheon+, tightly constraining an octopole remains a challenge. 
For this reason, 
despite its potential significance, 
we neglect the octopole contribution and further simplify the expression to 
\begin{equation}
    \label{eq: multipole_decel}
    \mathfrak{Q}(\boldsymbol{e}) \approx -1 - \frac{1}{\mathfrak{H}^2(\boldsymbol{e})} \bigg( \overset{0}{\mathfrak{q}}+e^\mu \overset{1}{\mathfrak{q}}_\mu  \bigg{)}.
\end{equation}

Similarly, for the effective Hubble parameter $\mathfrak{H}$, we use a simplified version of Eq. \eqref{eq:effective hubble_multipole_expansion} {in which we neglect} 
the dipole term:
\begin{equation}
    \label{eq: simplified effective hubble}
    \mathfrak{H}(\boldsymbol{e}) \approx \frac{1}{3}\theta - e^{\mu}e^{\nu}\sigma_{\mu\nu}.
\end{equation}

Armed with a  simplified version of the general cosmographic parametrisation {of the luminosity distance}, we now move toward constraining {these} anisotropies in 
{observational} data.


\color{black}

\section{Observational data and methodology}
Here we detail our methods for constraining the anisotropies discussed in the previous section using supernova data. We introduce 
{the use of supernovae distances in constraining cosmology} in Section~\ref{sec: Sne}, the statistical method we use in Section~\ref{sec:chisq}, the data sets we use in Section~\ref{sec:P+}, the parametrisations we constrain in Section~\ref{sec:theory_sn}, and present our constraints themselves in Section~\ref{sec:obs_results}.

\subsection{Supernova distances}
\label{sec: Sne}

Type Ia supernovae (SNe) are a key part of the cosmic distance ladder {\citep[see, e.g.][for a review]{Goobar2011}}. 
When calibrated, they act as standard candles for {distance} measurements 
{via} the relation 
\begin{equation}
    \label{eq: Tripp formula}
    \mu \equiv m_B^* - M ,
    \end{equation}
where $\mu$ is the distance modulus, $m_B^*$ is the corrected apparent magnitude {of the SN} and $M$ is {its} absolute magnitude. In this work, we will use 
$m_B^*$ from the Pantheon+ data set, {which we introduce in Section~\ref{sec:P+} below.}
{We refer the reader to} \citet{scolnic2021pantheon+} for details of the corrections applied {to $m_B$ in the public data set}. 

The luminosity distance, $d_L$ {(in units of 10 parsec)}, is related to the distance modulus {via} 
\begin{equation}
    \mu = 5 \;\log \left(\frac{d_L}{10pc}\right),
\end{equation}

{which then allows us to constrain a chosen}
cosmological model {via an analytic expression for the luminosity distance {$d_L(z)$ using}
the {observed} redshift of the SNe}. 
{An option would be to use} the \lcdm\, distance-{redshift}
relation{, however, this} {would} require input knowledge of the 
parameters for each component of the total energy-density {of the Universe}. 
Alternatively, at low redshift we can free ourselves from these constraints by using the cosmographic approach and {thus adopting} a general expansion history, as we {outlined earlier} in  Section~\ref{sec:general_cosmog}.

{While} the parameters describing the anisotropies {in the luminosity distance} 
are not degenerate with the SN~Ia absolute $B$-band magnitude {(see Section~\ref{sec:theory_sn}), the monopole of the Hubble expansion does suffer this degeneracy. This is well-known in SN~Ia cosmology and is the reason that calibrators are required for local $H_0$ measurements \citep[see, e.g.][]{Riess_2022,Freedman:2019}}. In the fiducial analysis for constraining the anisotropies, we only use Hubble flow SNe~Ia (i.e. with $z \geq 0.023$), hence, we cannot constrain the monopole, $H_{\rm mono}$. However, {in attempt to assess} 
the impact of anisotropies on the 
{``Hubble tension'', in Section~\ref{sec:Htension} we perform an analysis including the calibrator sample of SNe~Ia in Pantheon+ and simultaneously constrain the monopole and quadrupole of the Hubble expansion.}

\subsection{Constrained $\chi^2$ method}
\label{sec:chisq}
{In this work, we will place constraints on cosmological parameters (isotropic and anisotropic)}
{
by assuming a $\chi ^2$ distribution; }
\begin{equation}
    \chi^2_{SN} = \Delta ^T C_{SN}^{-1} \Delta
\end{equation}
{where the residual vector is} $\Delta = m_{obs} - m_{th}$ {(and $\Delta^T$ is its transpose vector)}, 
and $C_{SN}$ is the covariance matrix. We use {\tt pymultinest} \citep{pymultinest}, a python wrapper of {\tt Multinest} \citep{Multinest_Feroz_Hobson_Bridges_2009} to derive the posterior {distribution} of the parameters.

\subsection{Pantheon+ Data}\label{sec:P+}

{We use the new} 
Pantheon+ \citep{scolnic2021pantheon+} data set, {a first analysis of which was presented}
in \citet{P+cosmology-2021} and \citet{Riess_2022}. 
This new Pantheon release has added {six} 
large {SNe} samples to the original data set \citep{pantheon_2018}, including 574 more light curves at $z<1$, as well as updated surveys due to a better understanding of the photometry. It contains 1701 light curves from 1550 {SNe} 
in the redshift range $0.001 \leq z \leq 2.26$. 
{In Figure~\ref{fig:P+_skymap} we show a skyplot of the directions of the Pantheon+ SNe in galactic coordinates, coloured according to their redshift in the heliocentric frame, $z_{\rm hel}$. The bottom panel shows a histogram of the number of SNe in the sample as a function of $z_{\rm hel}$. }
The sky-coverage is 
{close to} isotropic {for the low-redshift SNe, however, the higher-redshift SNe are more strongly clustered on the sky.} 
{From the histogram, we can} see that the data is dominated by the low redshift {SNe}. 

{As briefly mentioned in Section~\ref{sec:cosmo_distances}, the cosmographic expansion of luminosity distance with redshift as a parameter is strictly only convergent for $z<1$ \citep{Cattoen:2007}. Since the Pantheon+ sample contains objects out to $z=2.26$, an upper-limit redshift cut might be necessary to ensure robust results. }
For our fiducial analysis, we use the same redshift cuts as {was used for the cosmographic fits for $H_0$ and $q_0$ in} \citet{Riess_2022}, 
{namely}, $0.023 \leq z \leq 0.8$. This {reduces our fiducial SNe} sample to 1341 light curves.

\begin{figure}
    \centering
\includegraphics[width=\columnwidth]{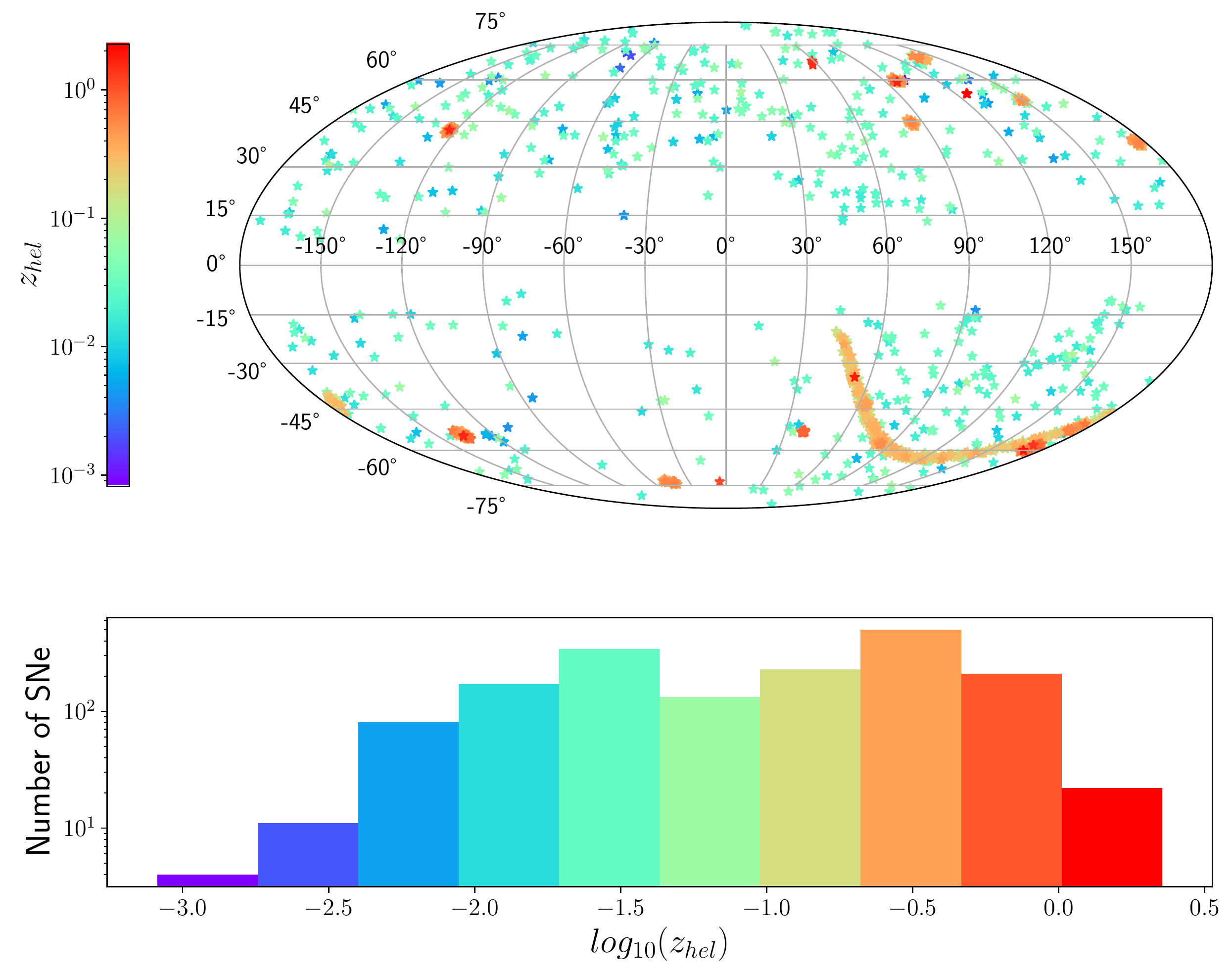}
    \caption{Skymap plot for the Pantheon+ dataset. Each star shows the location of a SNe on the sky, with colour showing heliocentric redshift. The lower panel shows the redshift distribution of the data.}
    \label{fig:P+_skymap}
\end{figure}

\subsection{Redshift Frames and Peculiar Velocity Effects}
\label{sec:zframes}

The choice of redshift frame has been shown to affect the strength of the 
{constrained} dipole in the deceleration parameter (e.g., \citetalias{dhawan2022quadrupole},\citet{Rubin_2020}). 
{Different ``redshift frames'' are defined by applying some kind of peculiar velocity corrections to the raw redshifts that we observe, with the goal of transforming the observations into a different frame of reference. } 
{Most commonly, three} redshift frames {are used}; the heliocentric frame (HEL), the CMB frame, and the Hubble diagram frame (HD). The heliocentric frame refers to redshift in the frame of the Sun, while the CMB frame redshifts have been corrected according to a 
single pointwise boost of our observations into the rest frame of the CMB. This boost is performed using our velocity inferred from the CMB dipole (assuming it is purely kinematic in nature). 
The HD frame redshifts are 
the CMB frame redshifts with peculiar velocity (PV) corrections applied to the SNeIa. These corrections are calculated based on density maps of the local Universe and linear perturbation theory, and estimate the PV of each SNe with respect to the CMB rest frame. 
For discussion on the effect of cosmological reference frames we refer the reader to \citet{Calcino:2016jpu}. In Section~\ref{sec:dipconstraints} and \ref{sec:quadconstraints}, we 
study the effect of varying the redshift frame on our resulting constraints.

\subsection{Anisotropic distances for supernovae}
\label{sec:theory_sn}

{Here we discuss our method to constrain the dominant multipoles we identified} 
in Section~\ref{sec:sim_results} 
in the Pantheon+ data set.

{For the effective deceleration parameter, as discussed in Section~\ref{sec:sim_results}, we only constrain its dipole anisotropy.}
We re-parameterise Eq. \eqref{eq: approx q} in a manner similar to \citetalias{dhawan2022quadrupole};
\begin{equation}\label{eq:q_dip}
     \mathfrak{Q}(\boldsymbol{e}) = q_{\rm mono} + q_{\rm dip}(\boldsymbol{e}) \,\mathcal{F}(z,S_{\rm dip}),     
\end{equation}
where $\mathcal{F}(z,S_{\rm dip})$ is the scale dependence of the dipole and 
$q_{\rm dip}=\bf{q_{\rm dip}}\cdot \boldsymbol{e}$, where $\boldsymbol{e}$ 
is the direction of the supernova {and ${\bf q_{\rm dip}}$ is the dipole vector}}. In order to quantify the strength of the dipole, we define the amplitude of the dipole at a given redshift as follows:
\begin{equation}
    A_d = \Big|\Big| \frac{ q_{\rm dip}({\bf e})}{q_{\rm mono}} \Big{|\Big|} \,\mathcal{F}_(z,S_{\rm dip}) .\\
    \label{eq:dipamp}
\end{equation}
{To further reduce the degrees of freedom and improve our fits,} we also assume {this dipole is aligned with} the CMB dipole. {Such an alignment might be expected based on the results of \citet{Heinesen_Macpherson_2021_Hubble_flow}, where the authors found that the dipole in the effective deceleration parameter should be aligned with local density contrasts near the observer. Further, \citetalias{dhawan2022quadrupole} and \citet{Sarkar2019} both tested the best-fit direction of this dipole and found it to coincide with the direction of the CMB dipole. Thus, we proceed by fixing the direction of ${\bf q}_{\rm dip}$ to be $(l,b) = (264.021, 48.523)$\textdegree\ as found in \citet{aghanim2020planck}.} 

The {low-redshift} anisotropic effects we are 
{interested in} are generated by the local clustering of matter, {which leads} to an anisotropic expansion of the {local space-time}. 
{We expect such effects to decay as we move further away from the observer, i.e., to higher redshift.}
For these reasons, {we choose to constrain} an exponentially decaying anisotropy \citep[see][for fits using other forms of $\mathcal{F}$]{Sarkar2019}. 
The exponential decay function {we use here} is 
\begin{equation}\label{eq:decay_function}
    \mathcal{F}(z, S) = {\rm exp}\left(-\frac{z}{S}\right),   
\end{equation}
{where we use} a uniform prior for the decay {scale}, 
$S$, in line with \citet{rahman2021new} and \citetalias{dhawan2022quadrupole}.

Other works have {similarly constrained a dipole in the deceleration parameter}
while 
{assuming} an isotropic $H_0$.
However, as we see from Eq.~\eqref{eq: deceleration_multipole}, $\mathfrak{Q}_0$ is dependent on $\mathfrak{H}_0$. 
In this work, we incorporate {the} 
{dominant quadrupole in} the Hubble parameter. Specifically, 
we re-parameterise Eq.~\eqref{eq: simplified effective hubble} as
\begin{equation}
\begin{aligned}\label{eq:h_quad}
    \mathfrak{H}(\boldsymbol{e}) &= H_{\rm mono} + H_{\rm quad}(\boldsymbol{e})\, \mathcal{F}(z,S_q) \\
    &= H_{\rm mono} \bigg\{1 + \bigg[\lambda_1 \cdot {\rm cos}^2\theta_1+ \lambda_2 \cdot {\rm cos}^2\theta_2 \\
    &\quad\quad - (\lambda_1 + \lambda_2) \cdot {\rm cos}^2\theta_3\bigg]\mathcal{F}(z,S_q) \bigg\},
\end{aligned}
\end{equation}
where {
$H_{\rm quad}(\boldsymbol{e})={\bf H_{\rm quad}}\cdot \boldsymbol{e}\, \boldsymbol{e}$ with
${\bf H_{\rm quad}}$ the quadrupole tensor,} $\lambda_1$ and $ \lambda_2$ {(and $\lambda_3 = \lambda_1 + \lambda_2 $)} are the eigenvalues of the {normalised} quadrupole moment tensor {${\bf H_{\rm quad}}/H_{\rm mono}$}, {the $\theta_i$ (for $i=1,2,3$)} are the angular separations between the location of the supernova and the quadrupole eigendirections, and $\mathcal{F}(z,S_q)$ is the quadrupole decay function. {We use the form of $\mathcal{F}$ given in Eq.~\eqref{eq:decay_function} but we ensure the decay scale $S_q$ is distinct from that of the dipole in the deceleration parameter.}

Due to the 
{additional} free parameters, it is difficult to constrain the direction of the quadrupole {to any useful accuracy with current data sets. Thus,} 
we assume 
{the quadrupole in the Hubble parameter is aligned} with 
{that} found in \citet{Parnovsky} {using the Revised Flat Galaxy Catalogue (RFGC) catalogue of 4236 galaxies.  
}
{In this work, the authors studied the} collective motion of local galaxies, fitting a dipole (bulk flow), a quadrupole (cosmic shear), and octupole component. They find eigenvectors {of the quadrupole} 
to have directions $(l, b)$ = (118, 85)\textdegree, (341, +4)\textdegree\ and (71, -4)\textdegree. This eigendirection was also used in the constraints on the quadrupole in the Hubble parameter in \citetalias{dhawan2022quadrupole}, where the authors found the constraints to be 
{{insensitive to} the chosen direction. }

{
{To assess the overall amplitude of the quadrupole in the Hubble parameter, we calculate} 
the amplitude of the quadrupole component of $\mathfrak{H}(\boldsymbol{e})$ in the same way as \citetalias{dhawan2022quadrupole}, and similar to the dipole amplotide defined in Eq. \ref{eq:dipamp}, {namely,} as the norm of the tensor ${\bf H_{\rm quad}}$ multiplied by the decay function $\mathcal{F}$, 
\begin{align}
    A_q &= || {\bf H_{\rm quad}} || \,\mathcal{F}_{\rm quad}(z,S) \\
    & = \sqrt{\lambda_1^2 + \lambda_2^2 + \left(\lambda_1 + \lambda_2\right)^2}\, \mathcal{F}_{\rm quad}(z,S_q), \label{eq:ampcalc}
\end{align}
for some redshift scale $z$. 
}

\section{Anisotropy in supernova data}
\label{sec:obs_results}
{In this section we present our constraints on local anisotropies in the Pantheon+ data set, motivated by those we identified in the numerical relativity simulations. 
In Section~\ref{sec:zframes} we discuss the redshift frames of the data we use, in Section~\ref{sec:dipconstraints} we present our constraints on the dipole in the deceleration parameter, and in Section~\ref{sec:quadconstraints} we present constraints on the quadrupole in the Hubble parameter. We also present constraints on the isotropic Hubble parameter, $H_0$, in Section~\ref{sec:Htension} and the resulting implications for the ``Hubble tension''.}


\begin{figure*}
    \centering
    \vspace{10mm}
    \includegraphics[width=.49\textwidth, ]
    {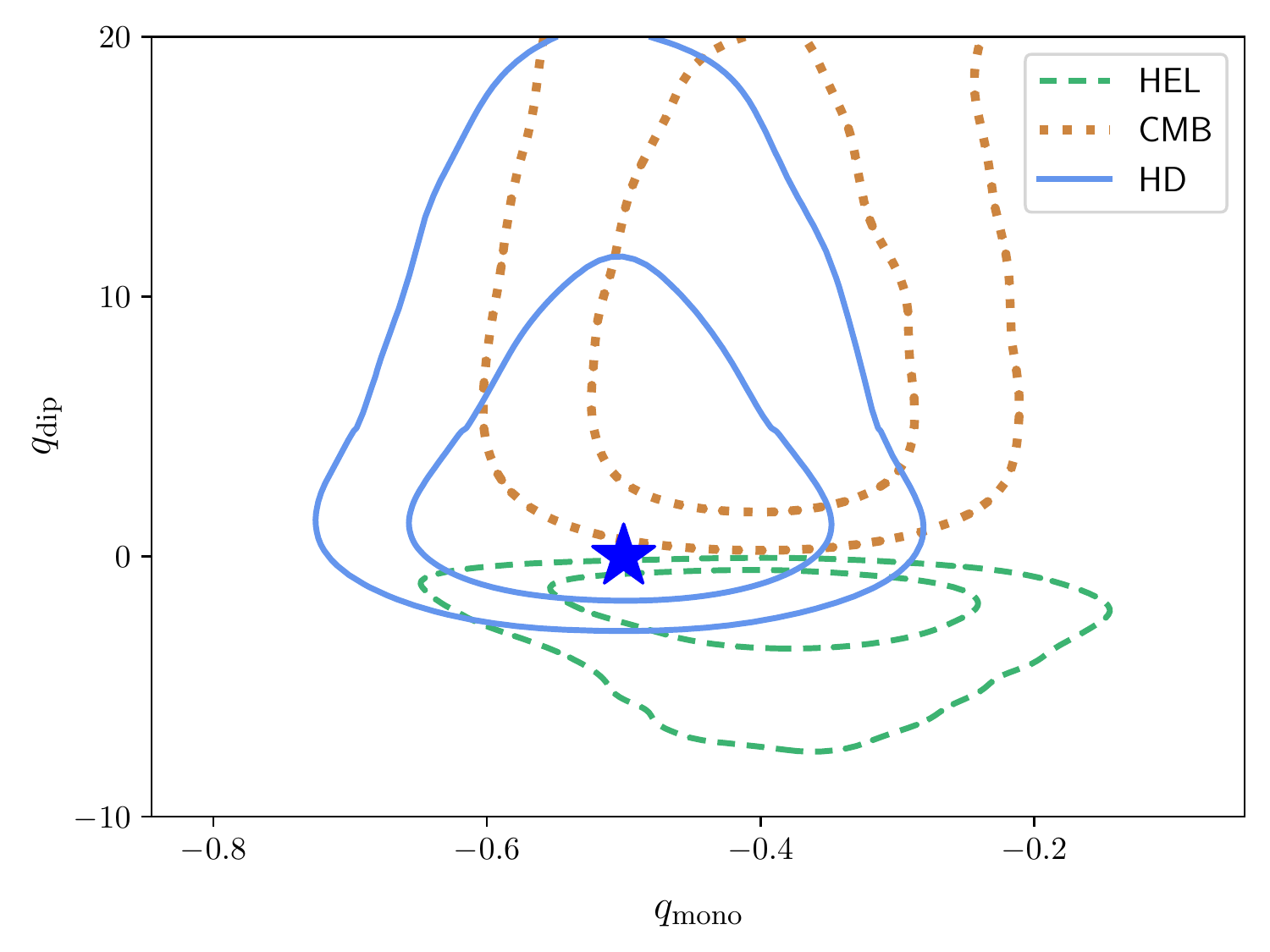}
    \includegraphics[width=.49\textwidth, ]{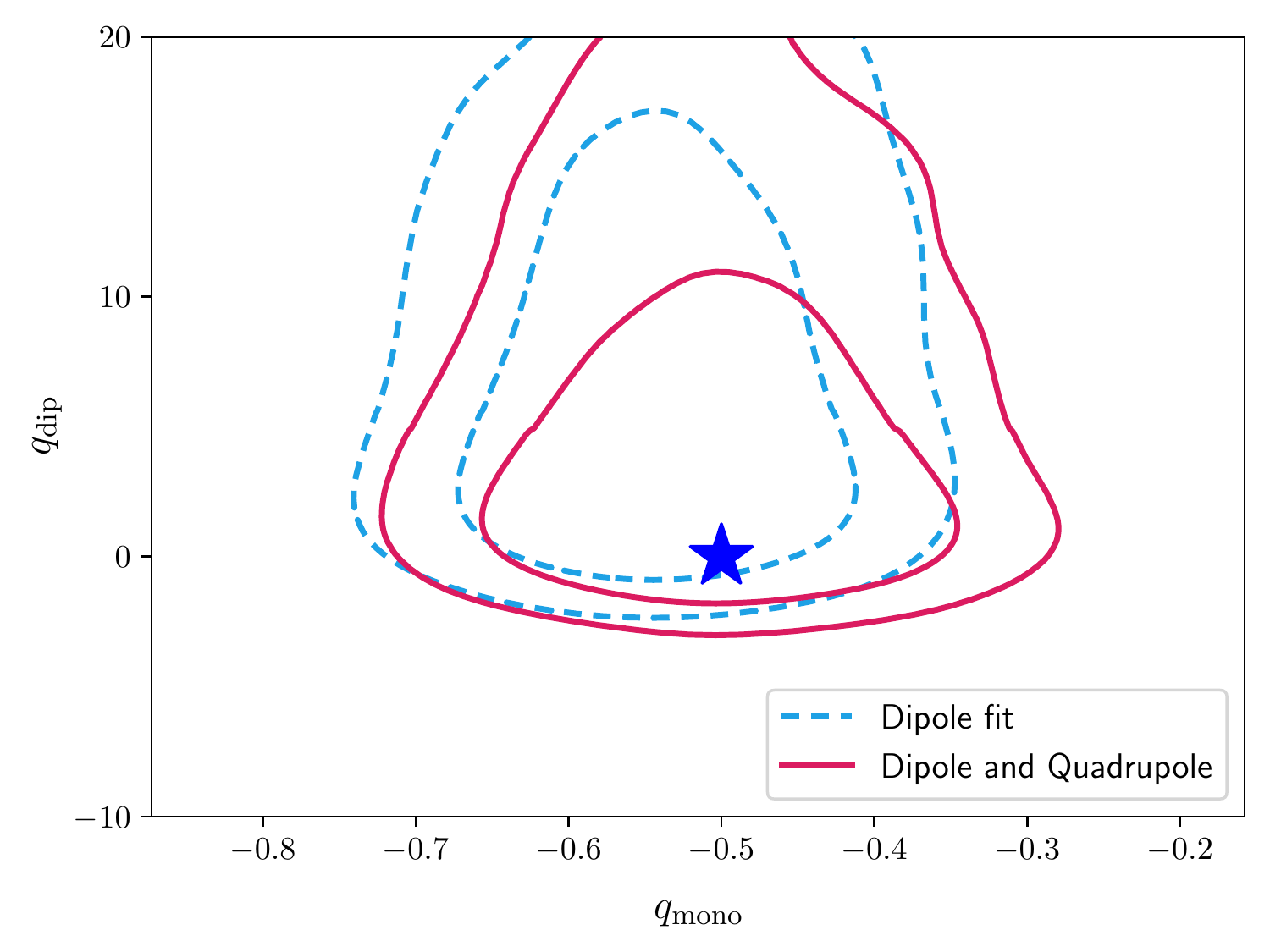}
    \caption{ Left panel: Results for the dipole in various redshift frames. The blue solid line shows our fiducial analysis, the Hubble diagram frame, the green dashed line shows the CMB frame and the yellow dotted line shows the heliocentric frame. Right panel:  Constraints on the dipole in the effective deceleration parameter when fitting only for a dipole (blue dashed) vs. also fitting for a quadrupole in the Hubble parameter with decay factor $S_q = 0.06/\ln (2)$(pink solid). The blue star marks isotropy. Contours represent the $1\sigma$ and $2\sigma$ –-constraints.
    }

    
    \label{fig:dipole}
\end{figure*}

\begin{figure*}
    \centering
    \begin{subfigure}{.49\textwidth}
        \includegraphics[width=\linewidth]{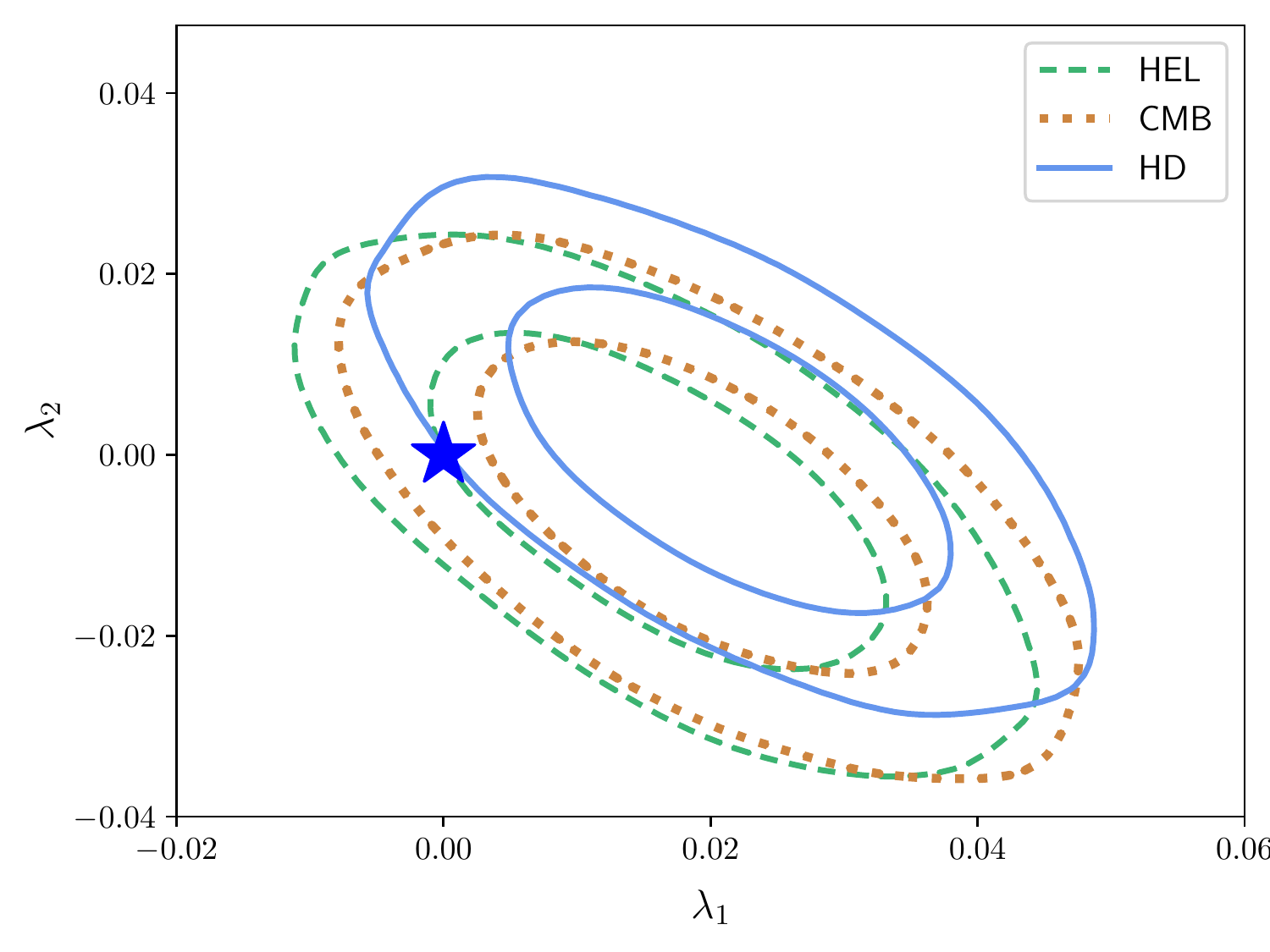}
    \end{subfigure}
    \begin{subfigure}{.49\textwidth}
        \includegraphics[width=\linewidth]{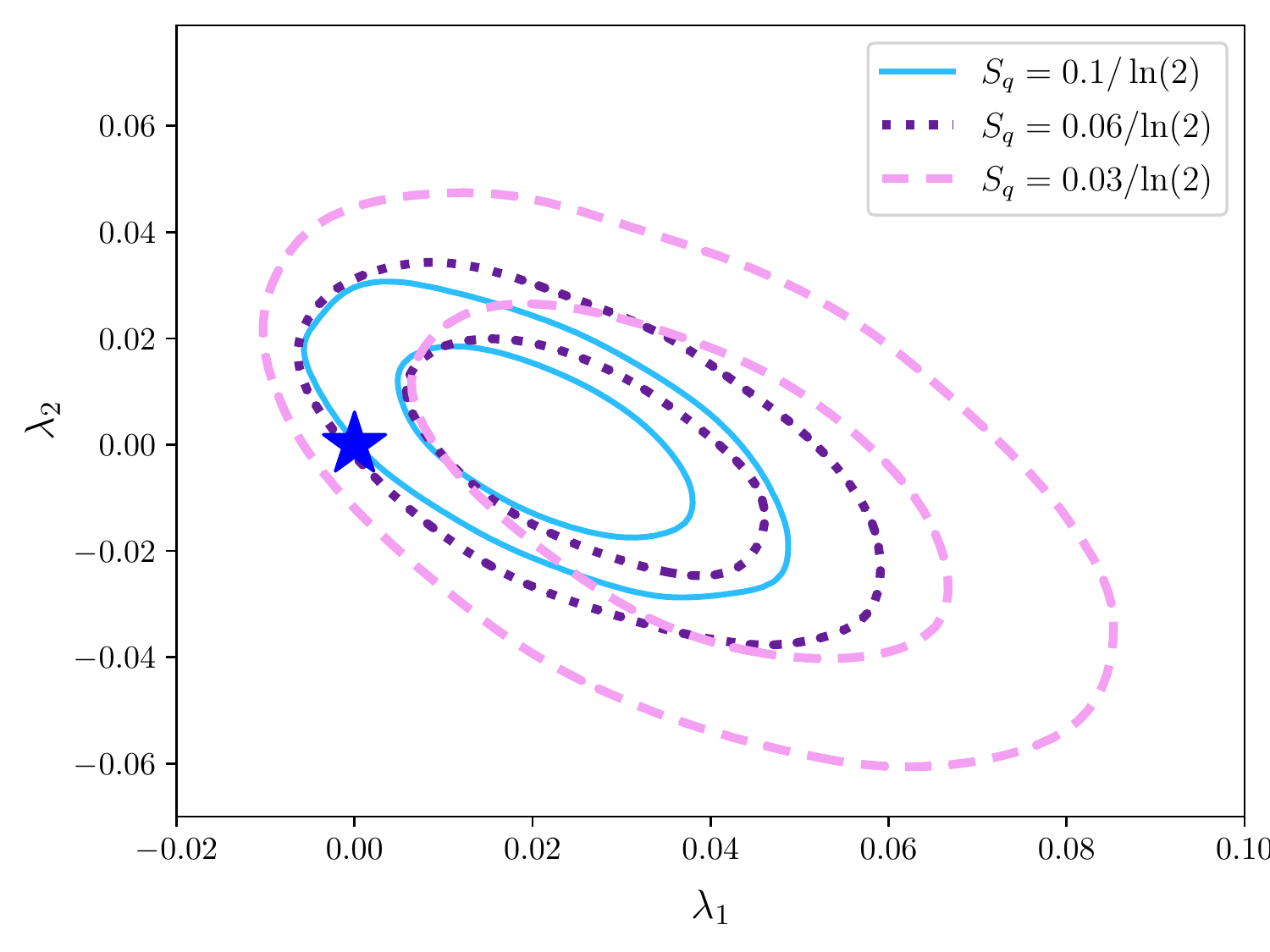}
    \end{subfigure}
    \caption{ Left panel: Constraints on the quadrupole eigenvalues in various redshift frames. The blue solid line shows our fiducial analysis, the Hubble diagram frame, the green dashed line shows the CMB frame and the yellow dotted line shows the heliocentric frame. The decay scale is set to $S_q = 0.1/\ln (2)$. The blue star marks isotropy. Contours represent the $1\sigma$ and $2\sigma$ –-constraints. {\bf Right panel:}Constrains on the eigenvalues of the quadrupole of the effective Hubble parameter when parameterised at fixed decay scales, $S_q$. The blue star marks isotropy. Contours represent the $1\sigma$ and $2\sigma$ constraints. All cases are inconsistent with isotropy.}
    \label{fig:quadrupole}
\end{figure*}

\subsection{Dipole of the deceleration parameter}\label{sec:dipconstraints}
We begin our analysis by fitting the dipole in the deceleration parameter of the form given in Eq.~\eqref{eq:q_dip}. 
{In this section, we will consider cases} both with and without an additional contribution from a quadrupole in the Hubble parameter. 
{As discussed in Section~\ref{sec:theory_sn},} we fix the direction of the dipole to coincide with the CMB dipole, as constrained by \citet{Planck_2018_III_High_freq}.
The dependence of the {resulting dipole amplitude} on this fixed direction has been explored in \citetalias{dhawan2022quadrupole} and \citet{Sarkar2019} and in both cases the CMB {dipole} direction was the best fit for the data sets. {We do not anticipate this dependence to differ in the case of the Pantheon+ data set.}

{First, we will constrain the dipole in the deceleration parameter in all three redshift frames introduced in Section~\ref{sec:zframes}, while simultaneously constraining a quadrupole in the Hubble parameter. In this section we present the constraints on $\mathfrak{Q}(e)$ only, and present the constraints on quadrupole parameters in the section \ref{sec:quadconstraints}. In the left panel of Figure~\ref{fig:dipole} we show constraints in the $q_{\rm dip}$-$q_{\rm mono}$ plane for heliocentric frame (dotted brown contours), CMB frame (dashed green contours), and HD frame (solid blue contours) redshifts. Contours represent the 1- and $2\sigma$ constraints for all fits. The blue star marks isotropy within $\Lambda$CDM, namely $q_{\rm mono} = -0.55$ and $q_{\rm dip} = 0$. 
We summarise the central values and 1$\sigma$ uncertainties of these constraints for all three frames in Table~\ref{tab:redshift_dipole}. 
When quoting $\sigma$ deviations from isotropy throughout this work, we do not assume a Gaussian distribution of our parameter constraints. {Instead, we use the inverse error function to convert a given percentile to a significance in multiples of $\sigma$ (in a similar manner to \citetalias{dhawan2022quadrupole}).}} 


In the HD redshift frame, we find consistency with isotropy at $<1\sigma$. 
However, in the CMB frame 
{we find} inconsistency with isotropy {{at $3.17 \sigma$. }
Our constraints for the CMB and HD frame redshifts are {roughly consistent} with \citetalias{dhawan2022quadrupole}, however, our heliocentric frame constraints are consistent with isotropy at $\lesssim 2\sigma$ while both \citetalias{dhawan2022quadrupole} and \citet{Sarkar2019} detect a dipole at $>2\sigma$. This difference can most likely be attributed to the addition of a quadrupole in the Hubble parameter in our analysis, since in both \citetalias{dhawan2022quadrupole} and \citet{Sarkar2019} the authors considered a dipole-only fit. It is perhaps the case that some of the anisotropy present in distances in the heliocentric frame is being absorbed into the quadrupolar anisotropy in $\mathfrak{H}(e)$ in our analysis.} 

\begin{table}
    \centering
    \begin{tabular}{c|c|c|c|}
    \hline
         Frame & $q_{\rm mono}$ & $q_{\rm dip}$ & Dipole significance \\
        \hline
        CMB & $-0.405\pm 0.089
        $& $9.6^{+4.0}_{-6.9}$ & 3.17 $\sigma$\\
                HD & $-0.503\pm 0.088
        $ & $ 4.5^{+1.9}_{-5.4}$ & - \\
                HEL & $ -0.391\pm 0.091
        $ &  $-2.36^{+1.6}_{-0.43} $ & $>2\sigma$\\
    \hline
    \end{tabular}
    \caption{{Summary of constraints on the dipole in the deceleration parameter for the three redshift frames used in the Pantheon+ data set. All results were obtained using the $\chi^2$ method and 
    significances were found using the highest posterior density interval and the error function, as in \citetalias{dhawan2022quadrupole}}}
    \label{tab:redshift_dipole}
\end{table}

{Next, we will compare our constraints on the dipole in the HD frame both} 
with and without a quadrupole contribution in the Hubble parameter. The right panel of Figure~\ref{fig:dipole}
shows our constraints assuming \textit{only} a dipole in the deceleration parameter {(dashed blue contours)}
and when also allowing for a quadrupole in $\mathfrak{H}$ with decay scale $S_q = 0.06/{\rm ln}(2)$ {(pink solid contours; see Section~\ref{sec:quadconstraints} below for constraints on the quadrupole and a discussion of decay scales)}.
Our constraints for the dipole only fit are 
are 
{$q_{\rm mono} = -0.541\pm 0.085$, $q_{\rm dip}=7.0^{+3.8}_{-7.2}$} {, to be compared with the HD case in Table~\ref{tab:redshift_dipole}, 
namely, we find a small shift to larger values of both $|q_{\rm mono}|$ and dipole magnitude for the dipole-only case. Although, this shift is{ $<1\sigma$} and so we conclude that the two cases are consistent with one another. } We also note that our results, for a similar selection of the redshift range, are consistent with a recent study by \citet{Sorrenti2022}.
From Eq. \ref{eq:dipamp}, using 1$\sigma$ constraints and a redshift of 0.035 and decay scale $S_{\rm dip} = 0.0367$, we find a {maximum} amplitude of $417\%$, and an amplitude of $345\%$ when using the median values. While not directly comparable due to having to select a redshift, this is in agreement with our simulation value of $110\%$  {For both cases, our constraints on the deceleration parameter} are also consistent with 
$\Lambda$CDM to within $1\sigma$. 


         

\color{black}
\subsection{Quadrupole in the Hubble parameter}\label{sec:quadconstraints}

\begin{table}
    \centering
    \begin{tabular}{c|c|c|c|}
    \hline
    $S_q$ &  $\lambda_1$ & $\lambda_2$ & $A_q$ \\
    \hline
     $ 0.1/{\rm ln}(2)$ & $ 0.021\pm{ 0.011}
$ & $ {3.15\times 10^{-5}}\pm 0.012$& $2.68\%$ \\
    
     $0.06/{\rm ln}(2)$ & $0.026 \pm {0.014}$&$-0.0021\pm { 0.014}$ & $2.88\%$ \\
 $ 0.03/{\rm ln}(2)$ &$ 0.037 \pm 0.02$& $ -0.0072\pm {0.022}$ & $2.85\%$ \\
    \hline
    \end{tabular}
    \caption{Constraints on the quadrupole in $H_0$ for the fiducial case in the HD frame 
    (while also fitting for a dipole in $\mathfrak{Q}(\boldsymbol{e})$). The {maximum allowed} amplitude {at $1\sigma$,} $A_q$, is calculated according to Eq.~\eqref{eq:ampcalc} at a redshift of $z = 0.035$, corresponding to scales of $\approx 100\,h^{-1}$ Mpc. }
    \label{tab:quad}
\end{table}

We constrain the effective Hubble parameter 
{using the form given in} Eq.~\eqref{eq:h_quad}. {Here, we will only} quote results from the joint fit of a dipole {in the deceleration parameter} and a quadrupole {in the Hubble parameter}. {We constrain the quadrupole eigenvalues $\lambda_1$ and $\lambda_2$ and initially fit with the decay scale $S_q$ as a free parameter in the analysis. However, we find $S_q$ to be largely unconstrained. Thus, we proceed to constrain the quadrupole eigenvalues for three values of the decay scale as used in \citetalias{dhawan2022quadrupole}. Namely, we choose }
$S_q=0.1/$ln(2), $S_q=0.06/$ln(2), and $S_q=0.03/$ln(2). This 
corresponds to a halving of the quadrupole amplitude at redshifts of $z=0.1, 0.06,$ and $0.03${, respectively}. 
{As mentioned in Section~\ref{sec:theory_sn}, to further} reduce the degrees of freedom we fix the quadrupole direction to that found by \citet{Parnovsky} {in the RFGC catalogue}. 
Specifically, {we fix the} eigenvectors {of the quadrupole} 
to {directions} $(l, b)$ = (118, 85)\textdegree, (341, +4)\textdegree\ and (71, -4)\textdegree. The {effect of varying the} quadrupole direction was explored in \citetalias{dhawan2022quadrupole}, {where the authors found} no significant improvement for different choices of eigenvectors.

As in \citetalias{dhawan2022quadrupole}, we also test the impact of redshift frame on the quadrupole constraints. 
{In the left panel of Figure~\ref{fig:quadrupole}, we show constraints on $\lambda_1$ and $\lambda_2$ for the heliocentric (HEL) frame (dotted brown contours), CMB frame (dashed green contours), and HD frame (solid blue contours) redshifts, where the blue star marks isotropy ($\lambda_1=\lambda_2=0$). These three constraints use a fixed }
decay scale of $S_q = 0.1/{\rm ln}(2)$. 
{As in} \citetalias{dhawan2022quadrupole}, we see {very} little shifting of the quadrupole posterior with redshift frame. We note however that {in the case of} the HD frame {redshifts, our constraints are}
inconsistent with isotropy at {$\sim 2\sigma$}{, while the heliocentric and CMB frame redshifts yield results consistent with isotropy at < 2$\sigma$. The CMB redshifts are calculated from the heliocentric redshifts using a single pointwise boost towards the CMB --- which is predominantly a dipolar correction. Thus,{we might expect \textit{some} shift in the} 
detected dipole in the deceleration parameter in the left panel of Figure~\ref{fig:dipole} (when moving from HEL to CMB redshifts).
{Although, we might not necessarily expect the CMB frame redshifts to still contain a dipole signal at $\sim 2\sigma$}. 

We would not expect this single boost to impact a quadrupole in the field of local SNe, which is what we find in the left panel of Figure~\ref{fig:quadrupole} (i.e., little to no shift moving from HEL to CMB redshifts). However, the next step to get from CMB to HD redshifts is to apply individual corrections to each SNe according to estimates of the local PV field. Such a correction is more extensive than a single boost, and thus 
in general should contain both dipole and quadrupole components {(as well as higher-order multipoles)}. 
This is what we find, namely, we see both a change in the dipole we detect (moving from CMB to HD redshifts in the left panel of Figure~\ref{fig:dipole}) \textit{and} a small, but noticeable, shift in the quadrupole (moving from CMB to HD redshifts in the left panel of Figure~\ref{fig:quadrupole}). This is perhaps surprising, as it implies including PV corrections, that is corrections of bulk flows, brings us further from the isotropic model by slightly pushing the quadrupole to higher values. However, we would expect the source  anisotropies to be due to bulk flows around an FLRW background, and so the corrections to have the opposite effect. 
For the rest of this section,} 
we take the fiducial case to be the HD frame {redshifts, since these are most commonly used in SNe analyses}.

{Next, we will assess the quadrupole eigenvalues when varying the fixed decay scale $S_q$.}
{In the} right panel of Figure \ref{fig:quadrupole}, {we show constraints on the two independent eigenvalues $\lambda_1$ and $\lambda_2$ for} 
a decay scale of $S_q=0.1/$ln(2) {with solid blue contours}, {a scale of} $S_q=0.06/$ln(2) with dotted purple contours, and $S_q=0.03/$ln(2) {with pink dashed contours}. The blue star again marks isotropy, i.e., $\lambda_1, \lambda_2$ = 0. 
For all cases {we consider}, we find a {quadrupole} signal {which is} inconsistent with isotropy at the $\sim 2\sigma$ level. {We calculate upper {($1\sigma$)} limits on the amplitude of the quadrupole{ using Eq~\ref{eq:ampcalc}} for all three cases, {and find
the largest amplitude 
{in the case of a decay} scale of $S_q=0.06/$ln(2), 
{namely, we place a limit of a $\lesssim$ { $2.88\%$} quadrupole strength at $z=0.035$.
In Table~\ref{tab:quad} we present our exact constraints together} with $1\sigma$ uncertainties.}} This is in agreement this with the strength we predicted from simulations in Section \ref{sec:sim_results}, {$\mathcal{R}_{\ell=2}(\mathfrak{H}) = 7.52 \times 10^{-3} = 0.752 \%$. }  

{While all contours are inconsistent with isotropy at {$\sim 2\sigma$,}
this deviation is only due to a non-zero $\lambda_1$ at 2$\sigma$, {while $\lambda_2$ remains consistent with zero at 1$\sigma$ for all three cases of fixed decay scale (and also for all three redshift frames in the left panel of Figure~\ref{fig:quadrupole}). }}
{Smaller values of $S_q$ imply a faster decay of $\mathcal{F}(z,S_q)$, and thus the quadrupole anisotropy being constrained is present at lower redshifts. We find that }
reducing the decay scale shifts the distribution of $\lambda_1$ {slightly} towards larger values {(though all distributions are consistent within 1$\sigma$)}. This is somewhat intuitive, 
{as} we expect these {local anisotropic} effects to 
{be larger closer to the observer, i.e., at lower redshifts}.
We also see {a widening of the constraints} 
for 
{progressively} smaller decay scales, which {again might} be expected because as we {sharpen the decline of $\mathcal{F}(z,S_q)$,}
we are {also effectively} constraining the anisotropy using less {SNe}.
Ideally, we would require a data set with more low-redshift objects to 
{more tightly} constrain these model{s with small decay scales}. 

{\citetalias{dhawan2022quadrupole} found the quadrupole in the Hubble parameter --- constrained using an identical method to that which we use here, though with the original Pantheon data set --- to be consistent with isotropy for all cases at $\sim 1\sigma$ using the same set of fixed decay scales as we show in the right panel of Figure~\ref{fig:quadrupole}. While our constraints are roughly consistent with those of \citetalias{dhawan2022quadrupole} to within $\sim 1\sigma$ for all three decay scales, we do see a shift to larger $\lambda_1$ values with the updated Pantheon+ data set, as well as a tightening of the contours.}

{Recently, \citet{Kalbouneh:2022tfw} defined a new observable based on a spherical harmonic decomposition of 
the observed Hubble expansion, correct to linear order in redshift. While the authors claim detection of a significant quadrupole in the \textit{Cosmicflows-3} all-sky galaxy catalogue \citep{CF3}, they report a null detection in the Pantheon sample \citep{pantheon_2018} as was found in \citetalias{dhawan2022quadrupole}. Such an analysis using updated data sets, such as \textit{Cosmicflows-4} \citep{CF4} and Pantheon+, would provide a valuable comparison and potential validation of the significant quadrupole we find in this work. The methods and data sets used in \citet{Kalbouneh:2022tfw} differ from our work such that a direct comparison of the quadrupoles we find is not straightforward. However, in the next section we will naively compare to their quoted maximal variance of $H_0$ across the sky and its relevance to the ``Hubble tension''. 
}

\subsection{Implications for the Hubble tension}\label{sec:Htension}

{Current measurements of the local Hubble expansion within the FLRW model, $H_0$,} using the Pantheon+ data set now lie in 5$\sigma$ tension with \lcdm\ predictions based on measurements of CMB anisotropies \citep{Riess_2022,aghanim2020planck}. {This is commonly referred to as the ``Hubble tension'' and no one resolution is widely accepted. }
In this section, we explore the significance of our {results with respect to measurements of the Hubble parameter}. 

\subsubsection{Impact on the monopole}

First, we assess the impact of accounting for a quadrupole on
the inferred value of the monopole of the Hubble parameter, i.e., for a measurement of the local Hubble constant $H_0$.  
We found a quadrupolar variance of the Hubble parameter across the sky at 2$\sigma$ significance. 
For a catalogue of SNe with incomplete sky coverage, an inference of the isotropic Hubble parameter $H_0$ might be expected to be impacted by this anisotropy. {Such an effect could result in a locally higher value of $H_0$ if the catalogue preferentially samples directions of maximal quadrupole amplitude \citep[see also][for a discussion on this]{Macpherson_Heinesen_2021}.}

{As briefly mentioned in Section~\ref{sec: Sne}}, SNe~Ia alone cannot constrain the monopole in the Hubble parameter 
{due to} the degeneracy between $H_{\rm mono}$ and the absolute luminosity of the SN~Ia, $M$.
{Including the}
calibrator SNe~Ia --- which are also distributed as part of the SH0ES and Pantheon+ data release --- {allows us to break this degeneracy and thus constrain $H_{\rm mono}$}. This {calibrator sample} includes a total of 37 galaxies with distances measured using Cepheid variables, hosting a total of 42 SNe~Ia. 
{We will first} 
simultaneously constrain $H_{\rm mono}$, $M$, and the parameters for {the} quadrupole {in the Hubble parameter}, i.e. $\lambda_1$ and $\lambda_2$. 
{We show our} resulting constraints {on the monopoles of the deceleration and Hubble parameters} 
in Figure~\ref{fig:h0_tension}. 
We use the same three fixed decay scales as in Section~\ref{sec:quadconstraints}, and for comparison, we show a purely isotropic constraint on $H_{\rm mono}$ with black dashed contours {(i.e., a fit with fixed $\lambda_1=\lambda_2=0$)}. 
{
The central value of $H_{\rm mono}$ {is 
{consistent across} all anisotropic fits, and }
does not differ significantly from the isotropic case. {We find the} largest difference {between the central values of the anisotropic fits and} the 
isotropic $H_{\rm mono}$ {to be} {0.30 km s$^{-1}$ Mpc$^{-1}$ in the case of $S_q = 0.1/\ln(2)$, where $H_0 = 73.40 \pm 1.02 $}. 
{Thus, we conclude} that despite finding a 
{2$\sigma$ significant} quadrupole, {accounting for this anisotropy in an inference of $H_{\rm mono}$ does not shift the central value enough to account for the $\sim 5$ km s$^{-1}$ Mpc$^{-1}$ Hubble tension discrepancy. }
}

\subsubsection{Maximal sky variance of the Hubble parameter}

\begin{figure}
    \centering
    \includegraphics[width=.46\textwidth]{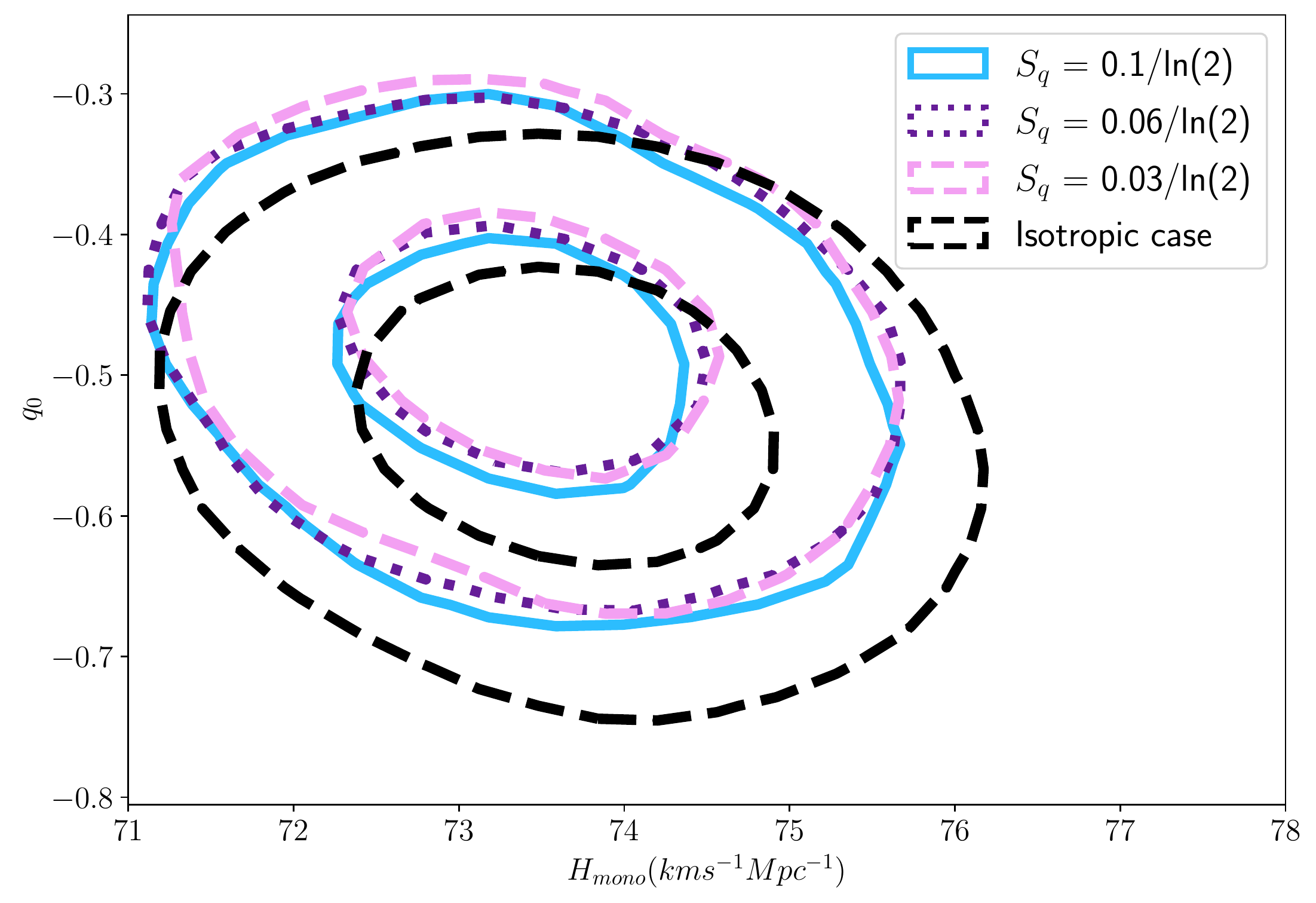}
    \caption{The constraints on the monopole of the Hubble parameter and the deceleration parameter with and without including anisotropy. Black dashed contours show an isotropic fit to the monopoles, and the coloured contours show constraints on the monopoles when accounting for anisotropy (for three models for the decay scale, as indicated in the legend). 
    }
    \label{fig:h0_tension}
\end{figure}

\citet{Kalbouneh:2022tfw} studied the maximal deviation in $H_0$ as measured using two populations of Pantheon SNe in antipodal directions on the sky. While the authors did not find a significant quadrupole feature in the Pantheon sample, they did find a dipolar feature in the distance modulus of these SNe \citep[which can most likely be attributed to anisotropy in the effective deceleration parameter, see][]{Heinesen_2021,Heinesen_Macpherson_2021_Hubble_flow}. For this sample, they find the maximal variance to be $\Delta H_0 = 2.4 \pm 1.1$ km s$^{-1}$ Mpc$^{-1}$ \textit{after} applying PV corrections (i.e. using HD frame redshifts). 
While here we constrained only a quadrupole anisotropy in the Hubble parameter, as is expected, we might still compare the maximal variance across the sky found by \citet{Kalbouneh:2022tfw} to our own given the constraints presented in Section~\ref{sec:quadconstraints}.
To do this we use the constraints on $\lambda_1$ and $\lambda_2$ given in Table~\ref{tab:quad} for the three decay scales studied, and calculate $\mathfrak{H}(\boldsymbol{e})$ using Eq.~\ref{eq:h_quad}. {We study two different cases of directions ($\boldsymbol{e}$) and redshifts ($z$ in the decay scale $\mathcal{F}(z,S_q)$) in calculating $\mathfrak{H}(\boldsymbol{e})$. First, we consider $\boldsymbol{e}$} 
given by the directions of \texttt{HEALPix} indices (with $N_{\rm side}=4$) and the separations $\theta_i$ are thus the sky separation between each \texttt{HEALPix} index and our fixed quadrupole direction. {In this first case, we consider a single redshift value of $z=0.035$ to quantify the variance in $\mathfrak{H}$ at $\sim 100 \, h^{-1}$ Mpc scales. 
Second, we consider $\boldsymbol{e}$ given by the directions of the Pantheon+ SNe and $\mathcal{F}(z,S_q)$ calculated using each SNe redshift, which lies in the range 0.023<$z$<0.8. This second case gives us a quantification of the variance in $\mathfrak{H}$ across the sky for the Pantheon+ SNe sample. In both cases, }
we then calculate $\Delta \mathfrak{H}(\boldsymbol{e})\equiv \mathfrak{H}(\boldsymbol{e})_{\rm max} - \mathfrak{H}(\boldsymbol{e})_{\rm min}$ for an assumed $H_{\rm mono} = 73.5$ km s$^{-1}$ Mpc$^{-1}$ (consistent with our fits in Figure~\ref{fig:h0_tension}). 

We show the variances we find in Table~\ref{tab:DeltaH0} {for both sample cases (single redshift or Pantheon+ redshift range) for all decay scale models we have constrained}. All show a $\sim 2--2.5$ km s$^{-1}$ Mpc$^{-1}$ sky-variance of the Hubble parameter, with upper limits closer to $\sim 4$ km s$^{-1}$ Mpc$^{-1}$. 
All of our results are consistent with the variance found in \citet{Kalbouneh:2022tfw} using the Pantheon sample{, although the authors here used a more restricted, low-$z$ sample of $0.01 < z < 0.05$ due to their first-order expression.}


 
\begin{table}
    \centering
    \begin{tabular}{c|c|c}
    \hline\hline
    $S_q$ & $z$ &  $\Delta \mathfrak{H}(\boldsymbol{e})$ (km/s/Mpc) \\ \hline\hline
     $ 0.1/{\rm ln}(2)$ & 0.035 &  $2.356^{+1.802}_{-1.185}$ \vspace{2mm} \\   
      & 0.023 < z < 0.8 & $2.329^{+1.693}_{-1.152}$ \\ \hline
     $0.06/{\rm ln}(2)$ & 0.035 & $2.377^{+2.010}_{-1.040} $ \vspace{2mm} \\
      & 0.023 < z < 0.8 & $2.406^{+1.999}_{-1.002}$ \\ \hline     
     $ 0.03/{\rm ln}(2)$ & 0.035 &  $2.122^{+1.853}_{-0.679}$ 
     \vspace{2mm} \\
           & 0.023 < z < 0.8 &  $2.471^{+1.948}_{-0.785}$ \\ 
    \hline\hline
    \end{tabular}
    \caption{Maximal sky variance of the anisotropic Hubble parameter, $\Delta \mathfrak{H}(\boldsymbol{e})$, for our best-fit quadrupole constraints given in Table~\ref{tab:quad}. All variances are calculated at a scale of $z=0.035$, corresponding approximately to $100\,h^{-1}$ Mpc. 
    }
    \label{tab:DeltaH0}
\end{table}




\section{Conclusions}
{
{The local Universe is highly inhomogeneous and anisotropic due to the presence of late-time nonlinear structures. }
This {naturally} leads to an anisotropic local expansion {of space-time}, which could impact {cosmological inferences which assume isotropy. 
In this work, {our goal was to} constrain theoretically-motivated anisotropies in 
low-redshift supernova data}. 

{We used} the {generalised cosmographic expansion of the luminosity distance from} 
\citet{Heinesen_2021} and the {simulation data} from \citet{Macpherson_Heinesen_2021} {to predict the dominant anisotropic signatures in nearby luminosity distances. To do this, we considered a set of two simulations, with different ``smoothing scales'', each containing 100 randomly-placed synthetic observers. Each observer has a full-sky distribution of the effective cosmological parameters defined in \citet{Heinesen_2021}, on which we performed a multipole expansion to determine the dominant contributions. }
We found 
{the} dipole and octopole {to be the dominant multipoles} in the {effective} deceleration parameter and {found the} quadrupole {to dominate the effective} Hubble parameter {for all cases we studied.} 
Within the simulation with smoothing scale of $100 \,h^{-1}$ Mpc, we found {the quadrupolar signal in the Hubble parameter has a} 
strength of $5.65 \times 10^{-3}\%$ with respect to the monopole {on average over all observers, while the} dipole in the deceleration parameter {has strength} $\sim 53 \%$ {on average}. 

{Next, we constrained these dominant anisotropies} using the {new} Pantheon+ {SNe} data set \citep{scolnic2021pantheon+}. {In the rest frame of the CMB, }
we found an {$ 3.17 \sigma$} significant dipole 
{with} magnitude {$q_{\rm dip}= 9.6^{+4.0}_{-6.9}$}. 
{When correcting SNe redshifts for their peculiar velocities (i.e. using HD frame redshifts), the significance is removed and we find consistency with \lcdm. \citep[see also,][]{Sorrenti2022}}
 {Interestingly, we found a {$1.96 \sigma$}} significant quadrupole in the Hubble parameter 
{even after applying all peculiar velocity corrections. }
{We place a new $1\sigma$ upper limit on the maximum amplitude of a quadrupole in the Hubble expansion of {$2.88\%$}. }

{Anisotropies in the Hubble expansion are of particular interest for the Hubble tension \citep{Macpherson_Heinesen_2021}. If the Hubble parameter varies depending on which direction we observe, studies assuming an isotropic Hubble expansion could be biased in their results. We performed an analysis in which we constrained the monopole of the Hubble parameter --- by also including calibrator SNe --- along with the anisotropic components, as shown in Fig. \ref{fig:h0_tension}. Allowing for such an anisotropic variance results in a monopole of the Hubble parameter of {$73.40 \pm 1.02$ km s$^{-1}$ Mpc$^{-1}$. }This corresponds to a maximum shift of {$\sim 0.30$ km s$^{-1}$ Mpc$^{-1}$} (for the cases considered here) with respect to the isotropic fit, and thus it is unlikely that such an anisotropic variance can account for the observed difference in local inferences of the Hubble parameter. 
}}

Finally, we note that our findings are specific to models within the cosmographic framework, and the effects we discuss arising in the simulation source purely from clustering effects. Therefore this work is not a general constraint on any potential anisotropy, especially e.g. anisotropic cosmological models such as e.g. \citet{Lavinto:Tardis, constantin2022spatially}.  

\section*{Acknowledgements}

JAC acknowledges support from the Institute of Astronomy Summer Internship Program at the University of Cambridge.
SD acknowledges support from the European Union's Horizon 2020 research and innovation programme Marie Skłodowska-Curie Individual Fellowship (grant agreement No. 890695), and a Junior Research Fellowship at Lucy Cavendish College, Cambridge. 
HJM appreciates support received by the Herchel Smith postdoctoral fellowship fund for the majority of this work. Support for the late stages of this work and HJM was provided by NASA through the NASA Hubble Fellowship grant HST-HF2-51514.001-A awarded by the Space Telescope Science Institute, which is operated by the Association of Universities for Research in Astronomy, Inc., for NASA, under contract NAS5-26555.

\section*{Data Availability}
All Python packages and the Pantheon+ data used in this work are publicly available. Specific code used in our analysis can be made available upon reasonable request to the corresponding author.




\bibliographystyle{mnras}
\bibliography{mnras_template.bib}



\appendix

\section{The Curvature and Jerk Parameters}
\label{sec: curv_jerk}
Here we present {our results on the level of anisotropy measured in} the effective curvature, $\mathfrak{R}$, and jerk, $\mathfrak{J}$, parameters {in the NR simulations presented in}
Section~\ref{sec:sims}. 
{The parameters} $\mathfrak{R}$ and $\mathfrak{J}$ are defined in Eq.~\eqref{eq:effective_curvature} and \ref{eq:effective_jerk} respectively. Their full multipole expansions {are given} in \citet{Heinesen_2021}.
{Here we perform the same multipole analysis on these parameters as we performed in Section~\ref{sec:sim_results} for the Hubble and deceleration parameters. }

\begin{figure}
    \includegraphics[width=\columnwidth]{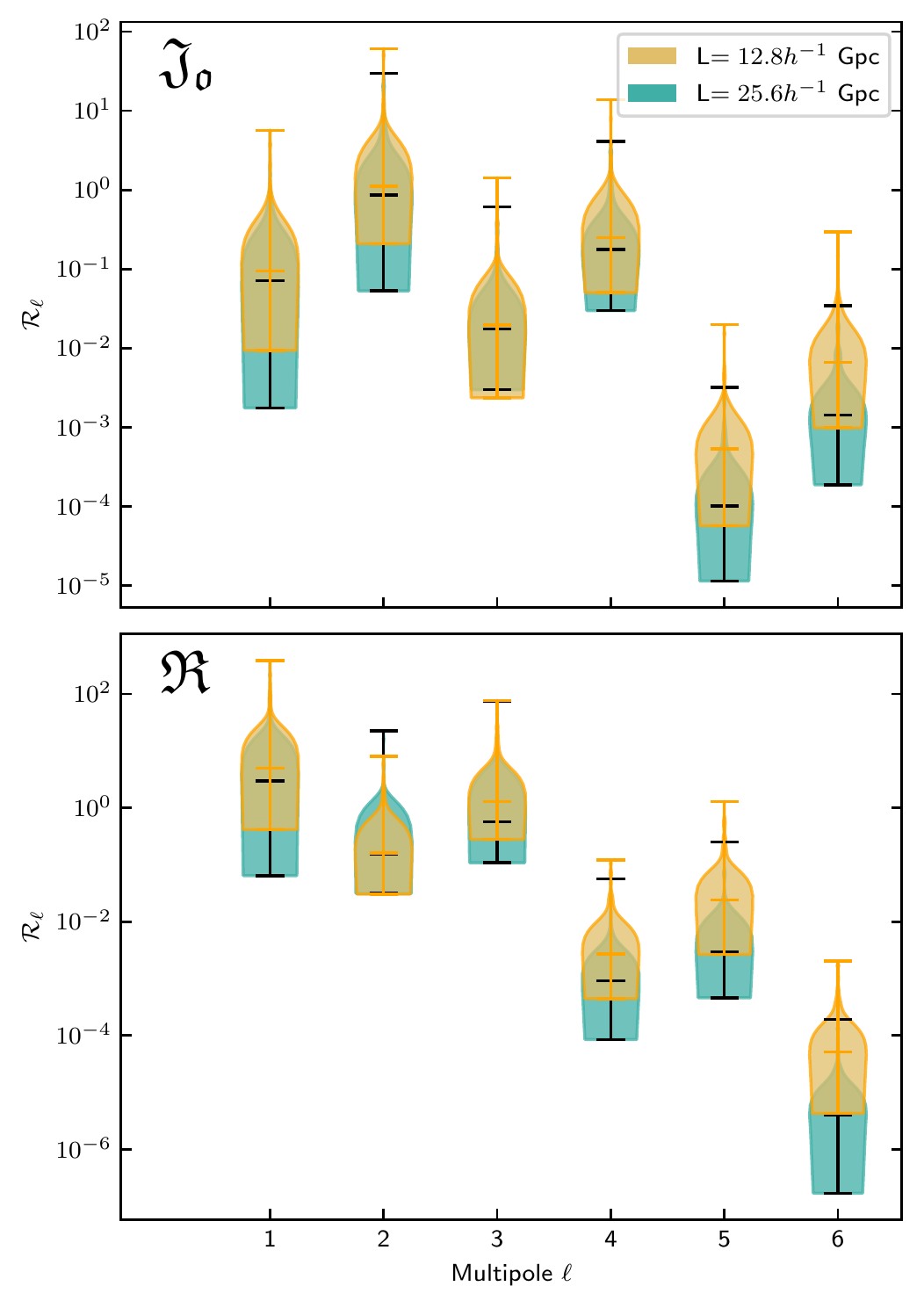}
    \caption{Relative ratios $\mathcal{R}_\ell$ for each multipole relative to the monopole for the effective jerk ($\mathfrak{J}_o$) and curvature ($\mathfrak{R}$) parameters. We show $\mathfrak{J}_o$ in the upper panel and in the bottom panel we show $\mathfrak{R}$. Shaded regions represent an empirical distribution of the data, calculated using kernel density estimation (KDE). Results for two simulations of length $12.8\,h^-1$ Gpc and $25.6\, h^-1$ Gpc are shown in yellow and blue, respectively. The upper limits, medians, and minimum values are marked on the plots with horizontal lines.}
    \label{fig:JandRviolin_plots}
\end{figure}

{Figure~\ref{fig:JandRviolin_plots} shows violin plots representing the strength of a multipole relative to the monopole, namely $\mathcal{R}_\ell$, as a function of the multipole number $\ell$. Yellow regions represent the distribution over 100 observers placed in the simulation with box length $L=12.8\,h^{-1}$Gpc, and green regions represent the same number of observers in the simulation with box length $L=25.6\,h^{-1}$Gpc. The top panel shows results for the effective jerk parameter and the bottom panel shows results for the effective curvature parameter. Horizontal bars in each distribution represent the maximum, mean, and minimum (top to bottom, respectively) value across the distribution of observers, and the width of each distribution represents the number of observers with that value of $\mathcal{R}_\ell$. }

As we also saw in Section~ \ref{sec:sim_results}, in general, the $L=12.8 h^{-1}$ Gpc simulation shows a trend of larger {amplitude} anisotropic effects. This is to be expected since this simulation has a smaller smoothing scale and thus higher density contrasts in general. We can also see that in both the {effective} jerk and curvature {parameters}, some higher-order {multipole} terms dominate over the isotropic terms {(i.e. they have $\mathcal{R}_{\ell} > 1$)}. {Specifically, {the median} 
${\mathcal{R}_{\ell=1}(\mathfrak{R}) =5.01}$ and ${\mathcal{R}_{\ell=3}(\mathfrak{R}) =1.30}$
.}
Similarly for $\mathfrak{J}_o$, we see the {median value of the ratio for the quadrupole is {$\mathcal{R}_{\ell=1}(\mathfrak{J}) = 1.12$,
.} }

\bsp	
\label{lastpage}
\end{document}